\documentclass[11pt]{article}
\usepackage{cite}
\usepackage{graphicx}
\usepackage{amsmath,amssymb}
\usepackage[body={6in,8.5in}]{geometry}
\usepackage{color}
\usepackage{url}
\usepackage{authblk}
\usepackage{caption}
\usepackage{graphicx}
\usepackage{caption}
\usepackage{cleveref}
\usepackage{array}
\usepackage{enumitem}
\usepackage{indentfirst}
\usepackage{graphicx}
\usepackage{caption}
\usepackage{subcaption}
\usepackage{array}
\usepackage{chngcntr}
\usepackage{hyperref}
\usepackage{bookmark}
\usepackage{booktabs}

\title {The Geometry of Large Tundra Lakes Observed in Historical Maps and Satellite Images}

\author[1]{Ivan Sudakov \thanks{isudakov1@udayton.edu}}
\author[2]{Almabrok Essa}
\author[3]{Luke Mander}
\author[2]{Ming Gong}
\author[1]{Tharanga Kariyawasam}
\affil[1]{Department of Physics, University of Dayton, Dayton, OH 45469, USA}
\affil[2]{Department of Electrical \& Computer Engineering, University of Dayton, Dayton, OH 45469, USA}
\affil[3]{School of Environment, Earth and Ecosystem Sciences, The Open University, Milton Keynes, MK7 6AA, UK}
\bigskip
\date{October, 2017}
\begin{document}
\maketitle



\section*{Abstract}
{\small The climate of the Arctic is warming rapidly and this is causing major changes to the cycling of carbon and the distribution of permafrost in this region. Tundra lakes are key components of the Arctic climate system because they represent a source of methane to the atmosphere. In this paper, we aim to analyze the geometry of the patterns formed by large (
$>0.8$ km$^2$) tundra lakes in the Russian High Arctic. We have studied images of tundra lakes in historical maps from the State Hydrological Institute, Russia (date 1977; scale $0.21166$ km/pixel) and in Landsat satellite images derived from the Google Earth Engine (G.E.E.; date 2016; scale $0.1503$ km/pixel). The G.E.E. is a cloud-based platform for planetary-scale geospatial analysis on over four decades of Landsat data. We developed an image-processing algorithm to segment these maps and images, measure the area and perimeter of each lake, and compute the fractal dimension of the lakes in the images we have studied. Our~results indicate that as lake size increases, their fractal dimension bifurcates. For lakes observed in historical maps, this bifurcation occurs among lakes larger than $100$ km$^2$ (fractal dimension $1.43$ to $1.87$). For~lakes observed in satellite images this bifurcation occurs among lakes larger than $\sim$100 km$^2$ (fractal dimension $1.31$ to $1.95$). Tundra lakes with a fractal dimension close to $2$ have a tendency to be self-similar with respect to their area--perimeter relationships. Area--perimeter measurements indicate that lakes with a length scale greater than $70$ km$^2$ are power-law distributed. Preliminary analysis of changes in lake size over time in paired lakes (lakes that were visually matched in both the historical map and the satellite imagery) indicate that some lakes in our study region have increased in size over time, whereas others have decreased in size over time. Lake size change during this 39-year time interval can be up to half the size of the lake as recorded in the historical map.}
\section{Introduction}
The Arctic is warming at approximately twice the rate of the rest of the globe \cite{duarte12,hinzman13}, and it is thought that climatic change in this region is proceeding so rapidly that the Arctic climate system is approaching a tipping point \cite{lenton08}. This is causing profound changes to the cycling of carbon in the Arctic and altering the spatial distribution of permafrost \cite{schu15}. Projected impacts of these changes range from local-scale alterations to the composition of vegetation in the Arctic, to socio-economic problems such as the reduced stability of buildings used for housing and industrial purposes \cite{hope}. 

Tundra~lakes are of particular concern in the context of Arctic climate change. This is because thawing permafrost and the resulting decomposition of previously frozen organic carbon significantly enhances the amount of methane emitted into the atmosphere under a warming climate \cite{Wik16}. Tundra~lakes, which change shape and increase in size as permafrost thaws, therefore represent a significant source of methane that provides a positive feedback to the atmosphere \cite{walt06,golubyatnikov13}, and they are critical elements of the Arctic climate system. Due to the central importance of tundra lakes in the Arctic carbon cycle, understanding their spatio-temporal dynamics is a key on-going scientific challenge \cite{and15}. 

Previous work on the spatio-temporal dynamics of lake-like patterns in the Arctic has shown that changes in their shape and patterning influence the structure of the Arctic climate feedback. For~example, studies conducted in Greenland and in the wider Arctic region have shown that surface melting during summer months has increased the abundance of supraglacial lakes on top of glaciers. These supraglacial lakes have lower albedo compared to ice and therefore absorb more of the sun’s energy, causing increased warming and potentially further melting on the ice sheet. Additionally, maps of the bathymetry of supraglacial lakes and streams, together with in situ measurements of their reflectance and depth, allowed Legleiter et al. \cite{leg13} to to measure the transient melt water flux through streams. Arctic sea ice melt ponds represent visible pools of collected meltwater on the sea ice surface, and these are capable of lowering albedo by altering light scattering properties of the ice surface \cite{hohenegger13}.
    
There are several potential sources of information on the spatio-temporal dynamic of tundra lakes, and each has its own advantages and disadvantages. It is possible to go into the field and measure the shapes and sizes of lakes manually. This has the potential to provide highly accurate information on lake geometry, but with limited spatial range, since it is not practical to survey hundreds of hectares in this fashion. Similarly, it is possible to analyze topographic maps that incorporate data on lake geometry. Such maps typically cover wide geographic areas, and can provide information on the character of the land surface from several decades ago, but they are not regularly updated owing to the time demands of gathering such data. Aerial photographs also provide a useful source of information on the distribution of elements on the land surface over wide geographic areas, but the cost of flying aircraft means that these data can have limited temporal resolution. Drone technology overcomes this barrier to a degree, but drones still require researchers to be present in the field in order to operate the craft. Satellite imagery provides a high volume of data on the land surface, and has provided valuable information on lake dynamics in Arctic permafrost regions \cite{kar13, nitze17,polishchuk15}. A challenge of working with satellite images is that the boundaries of objects of interest can be difficult to define, and image segmentation (the process of partitioning a digital image into multiple segments) can be difficult (for~example, see \cite{man17}). 
 
In this study, we broadly aim to analyze the geometry of the patterns formed by tundra lakes in the Russian High Arctic. This methodological paper represents a step towards understanding the spatio-temporal dynamics of tundra lakes in the Arctic, and builds on previous work using aerial photography and high-resolution satellite data to gather information on melt pond geometry and statistics (size and shape) \cite{hohenegger13, kim13}. We have chosen to investigate the potential for extracting geometrical information on tundra lakes from two sources: topographical maps and satellite images. Our specific objectives are: 
 \begin{enumerate} [leftmargin=*,labelsep=4.9mm]
\item{To develop image-processing routines that segment historical topographic maps and recent satellite images, and allow us to visualize tundra lakes in binary (black and white) images.}

\item {To calculate area--perimeter values for each of the individual lakes in our two sources of data.}

\item {To measure the geometrical properties of the tundra lakes in our two sources of data by calculating their fractal dimension.}
\end{enumerate}

\section{Materials and Methods}

\begin{figure}
\centering
   \includegraphics[width=.4\linewidth]{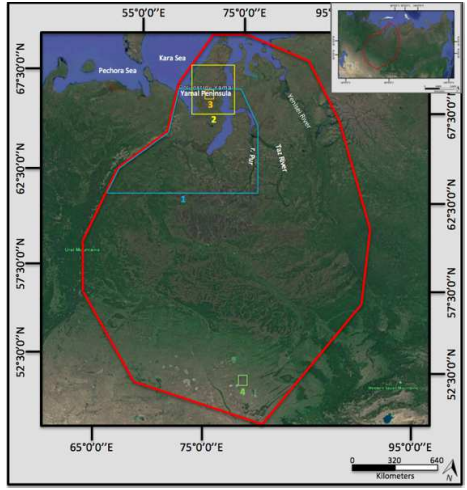}
   \caption{Overview map of the Western Siberia and with the study area. The box $1$ corresponds display area in Figure 2a; the box $2$ corresponds display area in {Figure} 3a; the box $3$ corresponds display area in {Figure} 3b; the box $4$ corresponds display area in {Figure} 9a-(1) and {Figure} 9b-(1).}
\label{fig-1}
\end{figure}

Our study was focussed on Western Siberia ($60^{\circ}{00}'{00}''$ Latitude and $75^{\circ}{00}'{00}''$~Longitude) (Figure~\ref{fig-1}), and, in this region, a long network of tundra lakes covers the West Siberian Plain~\cite{kar13, polishchuk13,kar14}. Our historical map \cite{h-map} was located in the State Hydrological Institute, Russia. This~map was created to display different types of wetland habitats, with the scale of 1:2,500,000, and the data underlying the map were collected from a series of field expeditions during the years preceding the map's date of publication (1969--1973). The map was created based on aerial photographs of the West Siberian Plain and field work. Detailed information about the analyses of aerial photographs obtained during this field work can be found in~\cite{iv76}.~The field work strategy included the setting of temporary stations in different areas of Western Siberia. Different institutions of the Academy of Sciences of the USSR collected the information from these stations during the study period and reported to the State Hydrological Institute (the field work coordinator). The summary of all field work was published in~\cite{rom70}. The map uses a variety of different colors and elements to display different types of wetlands, and the complexity of this color palette presents a challenge to the detection of lakes using computational vision. Our Landsat satellite images were gathered from the Google Earth Engine (G.E.E.) \cite{g-map} during 2016. Each image was taken at an eye height of $151.94$ km. The Google Earth Engine is a cloud-based and high-performance computing  platform for geospatial analysis contains remote sensing data from different dataset such as Landsat, MODIS, Sentinel and {ASTER} \cite{ge}.
This is a convenient platform for scientists working across different disciplines and is considered a useful tool for mathematicians, physicists and geologists, as well as climate and environmental scientists studying methane emissions from tundra lakes. The G.E.E. has been used for mapping of population dynamics, irrigation patterns (for example, see \cite{grl}), and the spatial patterns formed by vegetation in dryland ecosystems \cite{man17}.

\subsection{Detection of Lakes in Historical Maps}

\begin{figure}
	\begin{minipage}{0.5\textwidth}
	\centering
	\includegraphics[width=.65\linewidth]{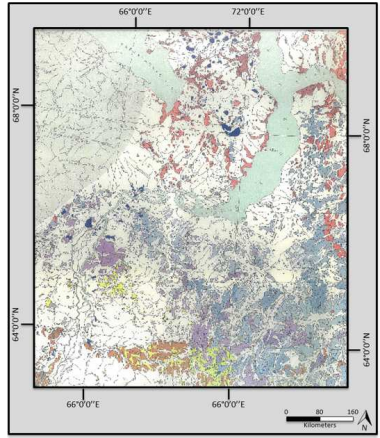}
	\subcaption{} 
	\end{minipage}\hfill
	\begin {minipage}{0.5\textwidth}
	\centering
	\includegraphics[width=.65\linewidth]{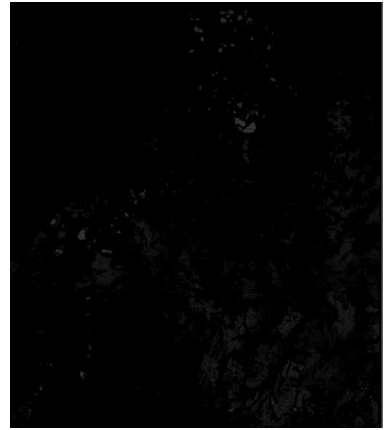}
	\subcaption{} 
\end{minipage}
\begin{minipage}{0.5\textwidth}
	\centering
	\includegraphics[width=.65\linewidth]{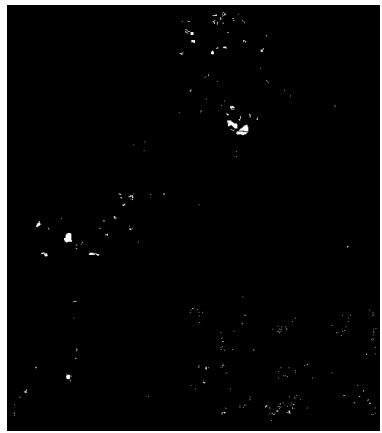}
	\subcaption{}
\end{minipage}\hfill
	\begin {minipage}{0.5\textwidth}
	\centering
	\includegraphics[width=.65\linewidth]{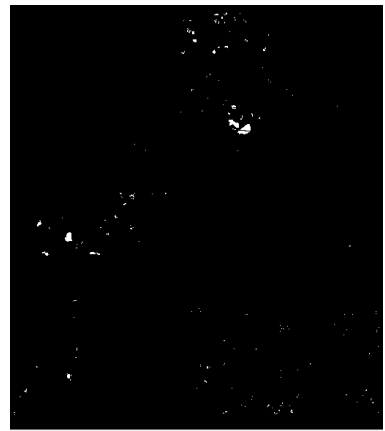}
	\subcaption{}
\end{minipage}
	\caption{Image processing steps taken to transform an image of a historical map into a binary image. (\textbf{a}) input image; (\textbf{b}) output image after applying color based segmentation; (\textbf{c}) binary image before applying line removal strategy; (\textbf{d}) binary image after applying line removal strategy.}
	\label{fig-2}
	  \end{figure}

In order to detect lakes in historical maps, the digital images of these maps were segmented based on their color intensity values. Figure \ref{fig-2}a shows a part of that historical maps. Firstly, each image was thresholded based on dark blue intensity values (Figure \ref{fig-2}b). Mathematically, given an input image $I_{RGB}(x,y)$, we extracted the blue channel $I_B(x,y)$  and then subtracted the gray-level $I_{gray}(x,y)$ image of the $I_B(x,y)$, which was done by

\begin{equation}
I_{blue}(x,y) = I_B(x,y) - I_{gray}(x,y),  
\end{equation}  
\noindent
where $I_{blue}(x,y)$  is the output image, and is the same size as the input image. This output image was converted to a binary (black and white) image in order to find the center of each lake (Figure \ref{fig-2}c). Topographical maps are characterized by many colors and lines, which are confounding in the context of image processing. To tackle this, we created a flat linear structuring element with the pixel of interest (the pixel being processed) located at its center. This flat linear structuring element was a matrix that contains zeros and ones. The pixels with values of $1$ define the neighborhood of pixels (structured as lines in the shape) that were included in the processing. The elements of the matrix with values of $0$ were not included. This structuring element is symmetric with respect to the neighborhood center, and there was a distance of $10$ pixels between the centers of the structuring element members at opposite ends of the line and an angle of $25$ degrees from the horizontal axis, which were each chosen by trial and error. We then subtracted that linear structure from the $I_{blue}(x,y)$ image (Figure \ref{fig-2}d). Our final image segmentation involved applying a region growing procedure to the images that were subjected to this line removal strategy. Since this region-growing step was common to both topographical maps and satellite images, we describe it separately below.

\subsection{Detection of Lakes in Google Earth Engine Images}\label{sec2.2}

In order to detect lakes in satellite images derived from the Google Earth Engine (Figure~\ref{fig-3}b), we~employed a~decision-making model using a support vector machine classifier (SVM) \cite{svm}, which~is a~supervised statistical and discriminative classification technique.~We trained the SVM using two~training samples of input images (size $7\times7$ pixels). The first training sample consisted of images representing lakes. The second training sample consisted of images representing the image background (not lakes). The vector information of each sample was extracted by representing each pixel $p$ as the Euclidean distance between its corresponding coordinates on the red, green, and blue channels and the origin, which was done by 

\begin{equation}
p(x,y) = \sqrt{(r-0)^2+(g-0)^2+(b-0)^2} = \sqrt{r^2+g^2+b^2},
\end{equation}
\noindent
where $r, g, b$ are the red, green, and blue channels of each sample ($7\times7$ sub-image) respectively of each pixel position $p(x,y)$.

\begin{figure}\label{fig-3}
\begin{minipage}{0.5\textwidth}
		\centering
		\includegraphics[width=.7\linewidth]{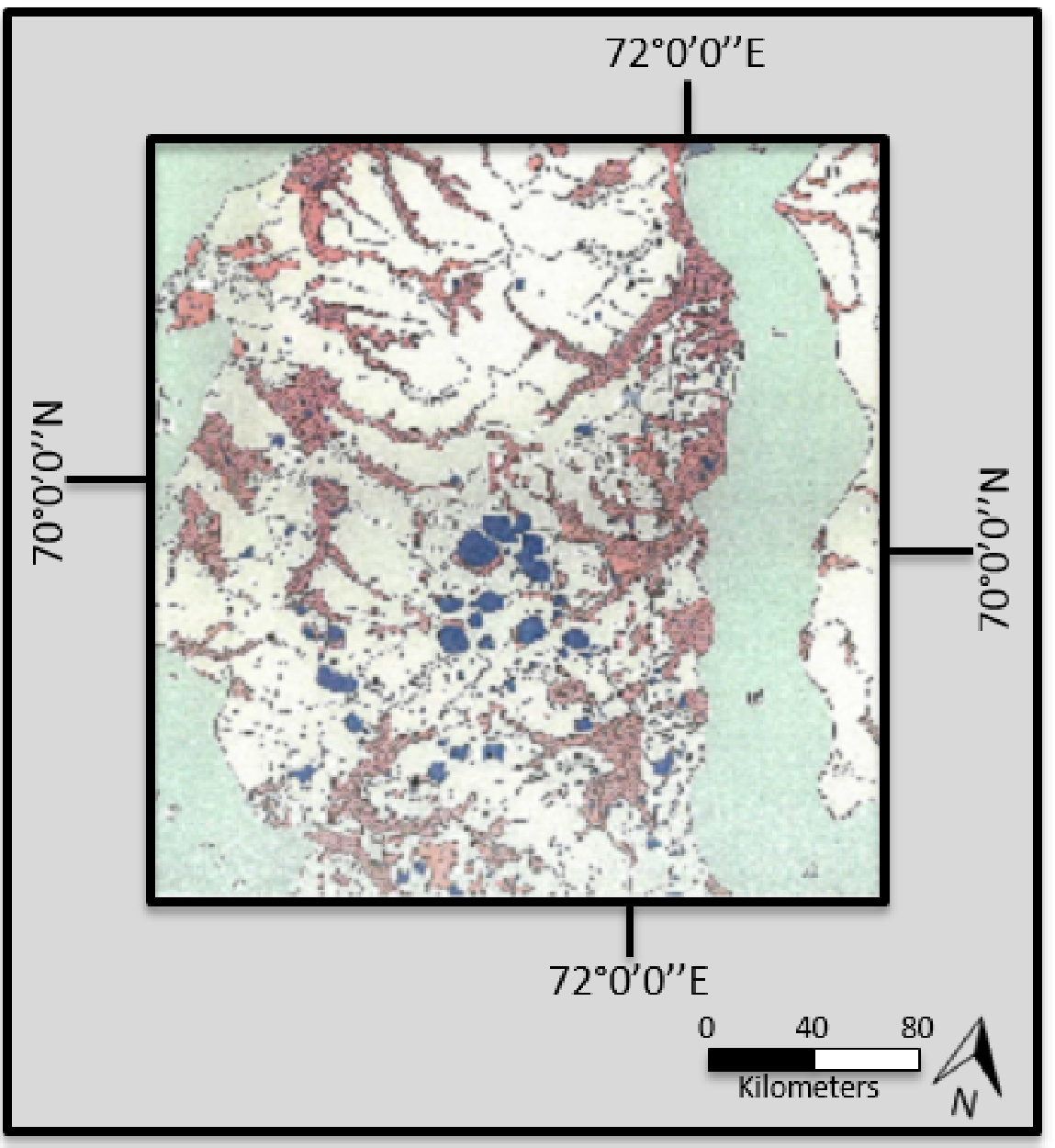}
		\subcaption{} 
	\end{minipage}\hfill
	\begin {minipage}{0.5\textwidth}
	\centering
	\includegraphics[width=.7\linewidth]{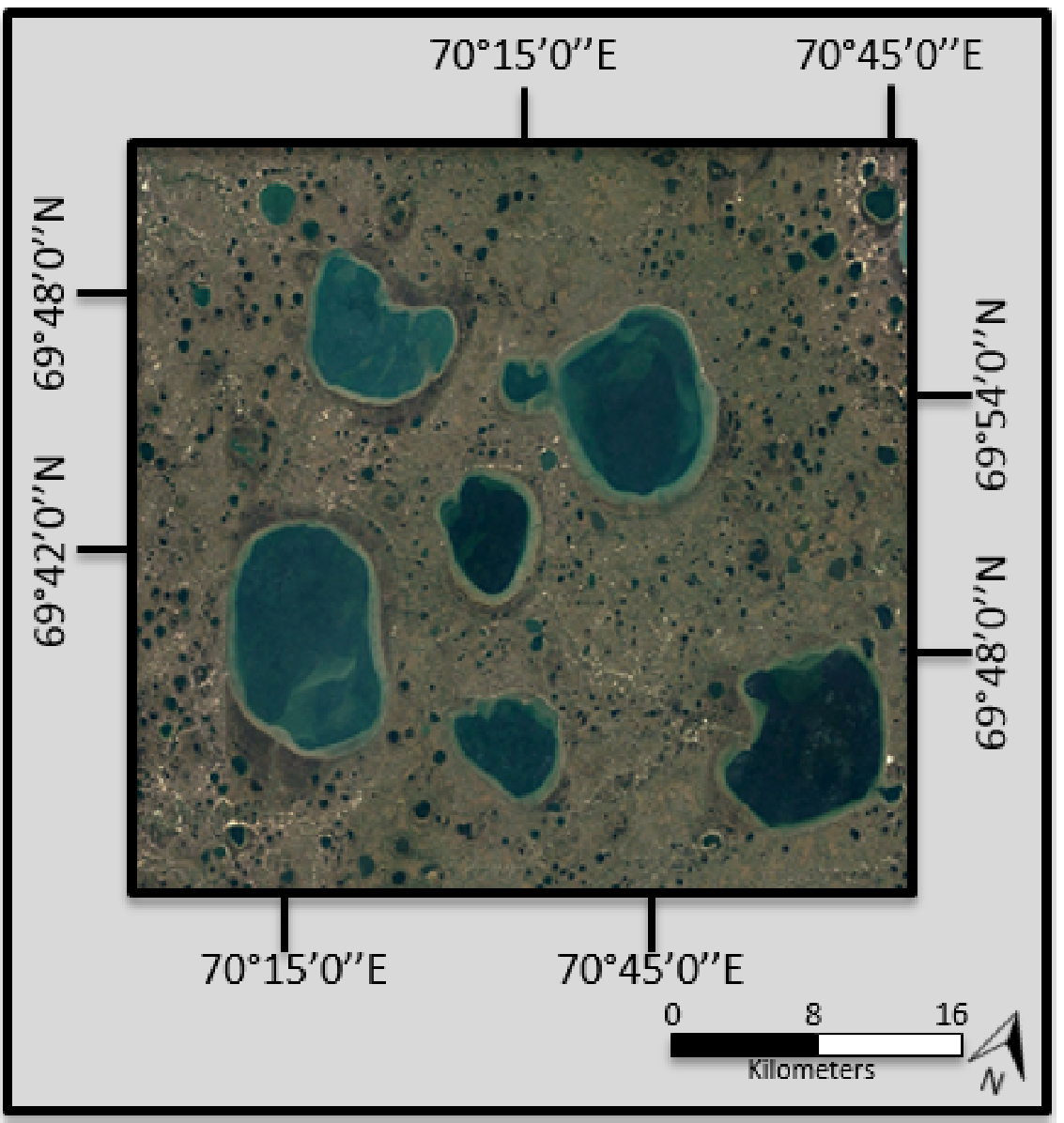}
	\subcaption{} 
\end{minipage}
\begin{minipage}{0.5\textwidth}
	\centering
	\includegraphics[width=.5\linewidth]{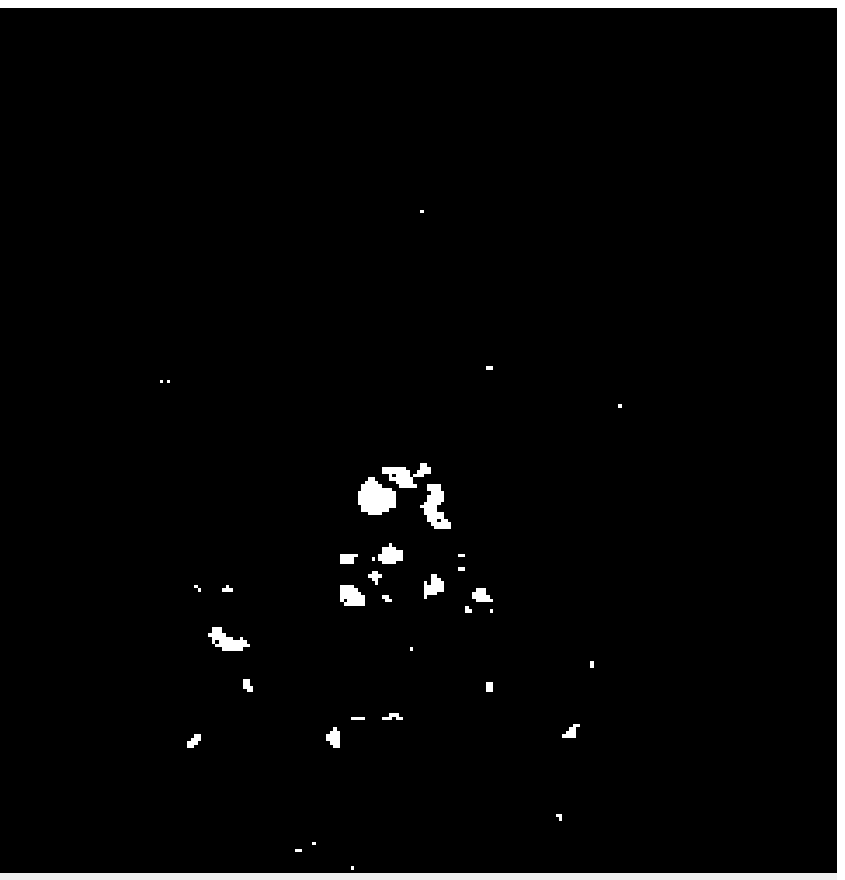}
	\subcaption{}
\end{minipage}\hfill
	\begin {minipage}{0.5\textwidth}
	\centering
	\includegraphics[width=.5\linewidth]{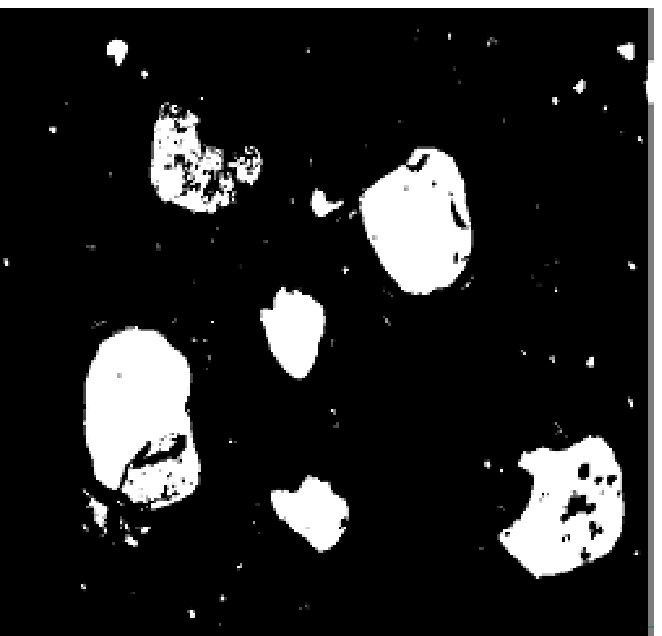}
	\subcaption{}
\end{minipage}
\begin{minipage}{0.5\textwidth}
	\centering
	\includegraphics[width=.5\linewidth]{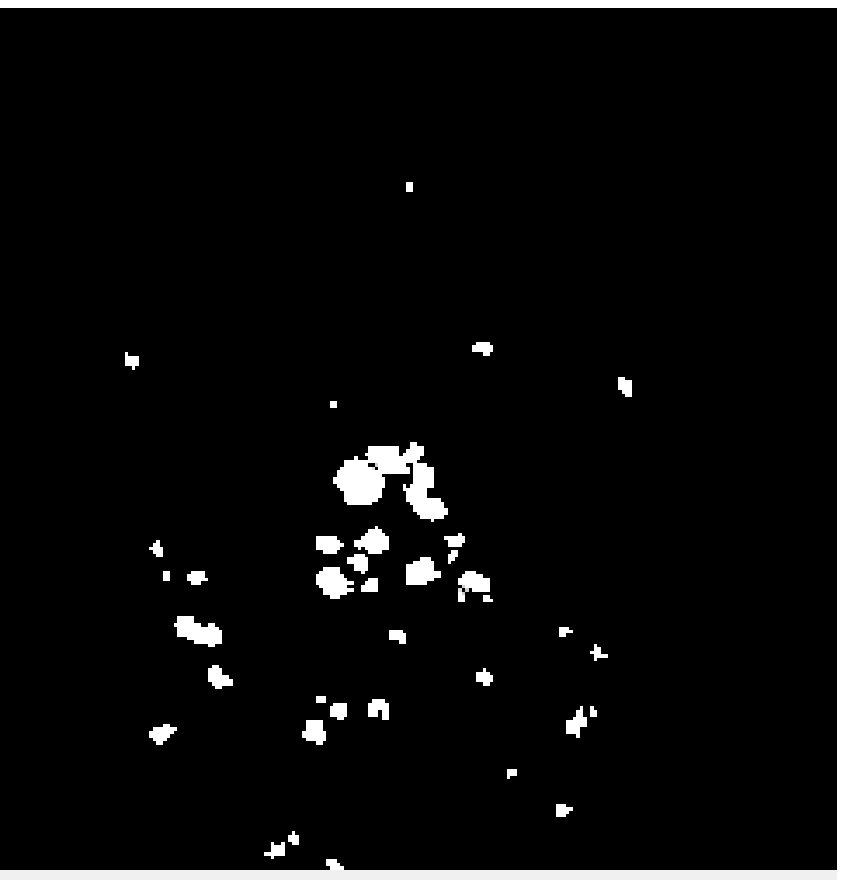}
	\subcaption{}
\end{minipage}\hfill
\begin{minipage}{0.5\textwidth}
	\centering
	\includegraphics[width=.5\linewidth]{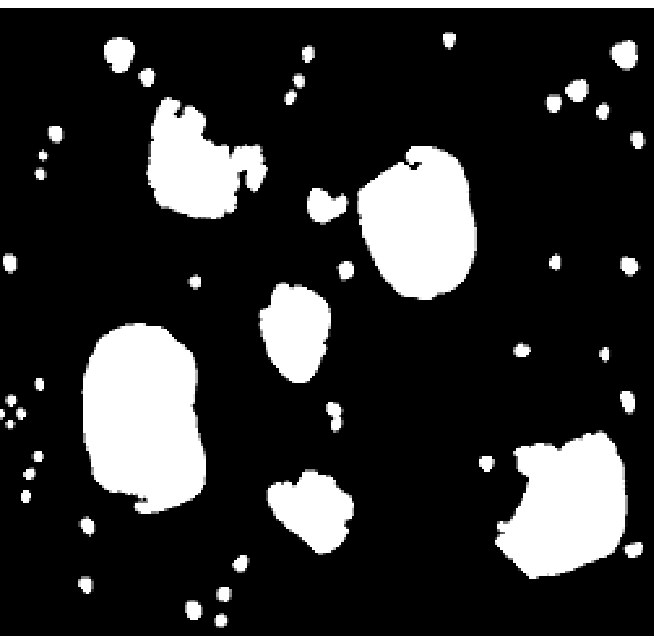}
	\subcaption{}
\end{minipage}

\caption{Region-growing image segmentation. Note the correspondence between lakes as seen in the input images and lake seen in binary images after region-growing. (\textbf{a}) historical input image; (\textbf{b}) Google Earth Engine (G.E.E.) input image; (\textbf{c}) binary image of historical before applying region growing strategy; (\textbf{d}) binary image G.E.E. before applying region growing strategy; (\textbf{e}) binary image of historical after applying region growing strategy; (\textbf{f}) binary image G.E.E. after applying region growing~strategy.}
\label{fig-3}      
\end{figure}

This vector information was fed to the support SVM classifier to build the SVM training model. Each input image was then divided to overlapping windows of size $7\times7$ pixels and its vector information was extracted in order to classify each window as a lake or not a lake.~Lakes were represented by white pixels and pixels that were not lakes were represented by black pixels (Figure~\ref{fig-3}d). Our final image segmentation involved applying a region growing procedure to the images that were subject to this line removal strategy.

 Region growing for segmentation of historical maps and Google Earth Engine images. We applied a standard region-growing segmentation to binary images of both historical maps (Figure \ref{fig-3}c) and satellite images (Figure \ref{fig-3}d). The pixel at the center of each lake was thought of as the pixel of interest. The difference between the intensity value of the pixel of interest and the intensity value of the region mean was calculated. If that difference was less than or equal to a predefined threshold, a new pixel was added to the region; otherwise, the process was stopped \cite{17, regiongrowingcode}. We experimented with a range of threshold values and found that $0.05$ yielded the clearest segmentation of lakes. This region-growing segmentation is shown in Figure \ref{fig-3}.  
  
\subsection{Calculating the Geometrical Properties of Lakes}  
We used standard connected component analysis to calculate the area and perimeter of each lake~\cite{18}.~These quantities were calculated in terms of pixels, and reported in metric units by multiplying the number of pixels with their corresponding pixel size. The pixel size for the historical map is $0.21166$ km/pixel, while the pixel size for the Google Earth Engine is $0.1503$ km/pixel.

The fractal theory introduced by Mandelbrot \cite{ mandelbrot77} can be used as a method to study partially correlated (over many scales) spatial phenomena that are not differentiable but are continuous. This theory helps quantify complex shapes or boundaries and relate them to underlying processes that may affect pattern complexity. For simple objects like circles and polygons, the perimeter $P$ scales as the square root of the area $A$. However, for complex planar regions with fractal curves as their boundaries,

\begin{equation}
P\sim \sqrt[D]{A},
\end{equation}
where the exponent $D$ is the fractal dimension of the boundary curve.

\begin{figure}
\centering
   \includegraphics[width=1.0\linewidth]{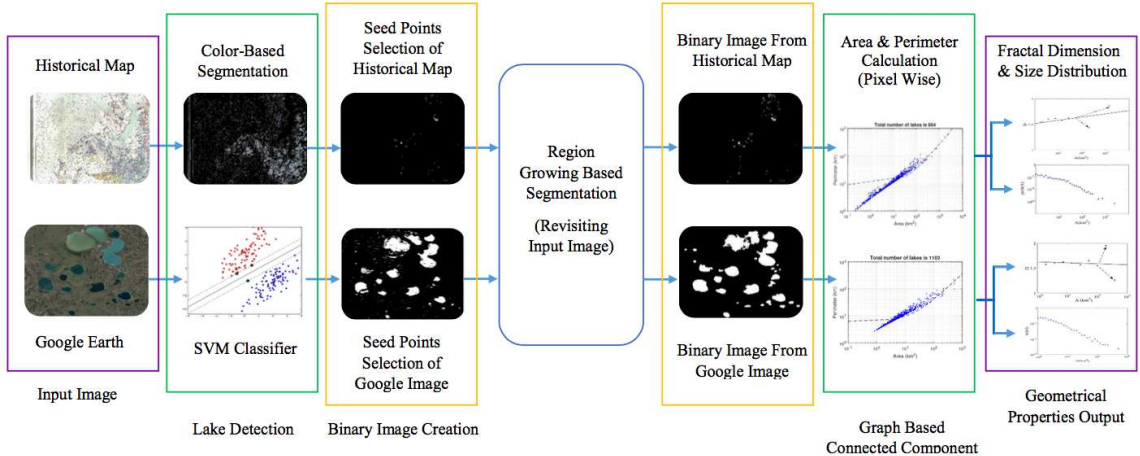}
   \caption{The algorithm of our study of tundra lake geometry.}
\label{fig-A3}
\end{figure}

We employed a method for computing the fractal dimension that is an extreme value analysis based on the lower edge of the area--perimeter data points. To do this, we took the convex hull of the data points in the ($A,P$)-plane, identified the lower edge, and computed its slope. This procedure should guarantee that fractal dimension does not decrease as area increases, and explains why there are so few data points that maintain the non-decreasing property of the fractal dimension. The method may be seen as a robust alternative to other methods of computing fractal dimensions, such as the lexicographic ordering method used for calculating melt pond fractal dimension \cite{hohenegger13}. We used a~scatter plot of perimeter--area obtained from the historical map and satellite images to compute fractal dimensions using MATLAB(R2015a, The MathWorks Inc., Natick, MA, USA). A summary of our method is shown in Figure \ref{fig-A3}.
\section{Results}
\vspace{-6pt}
\subsection{Geometrical Properties of Tundra Lakes in Historical Maps and Satellite Imagery}
\begin{figure}
	\begin{minipage}{0.5\textwidth}
	\centering
	\includegraphics[width=1.0\linewidth]{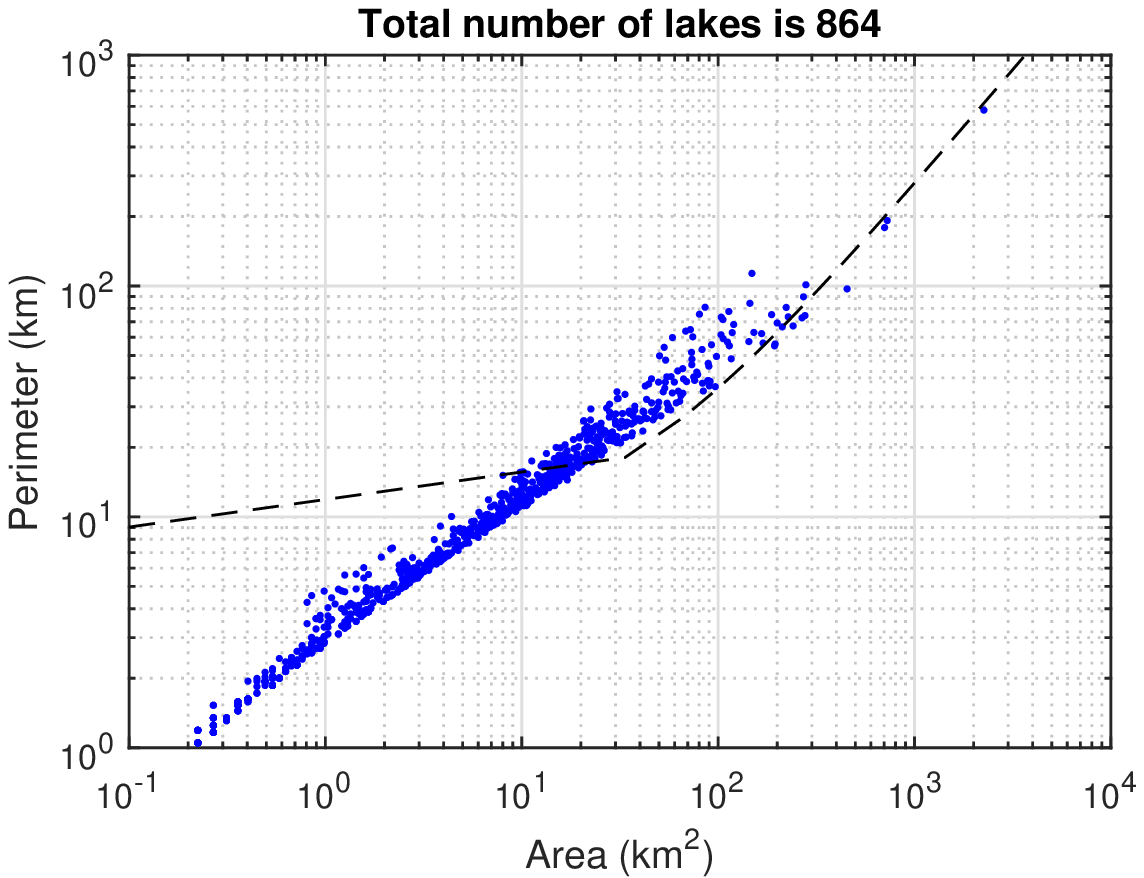}
	\subcaption{} 
	\end{minipage}\hfill
	\begin {minipage}{0.5\textwidth}
	\centering
	\includegraphics[width=1.0\linewidth]{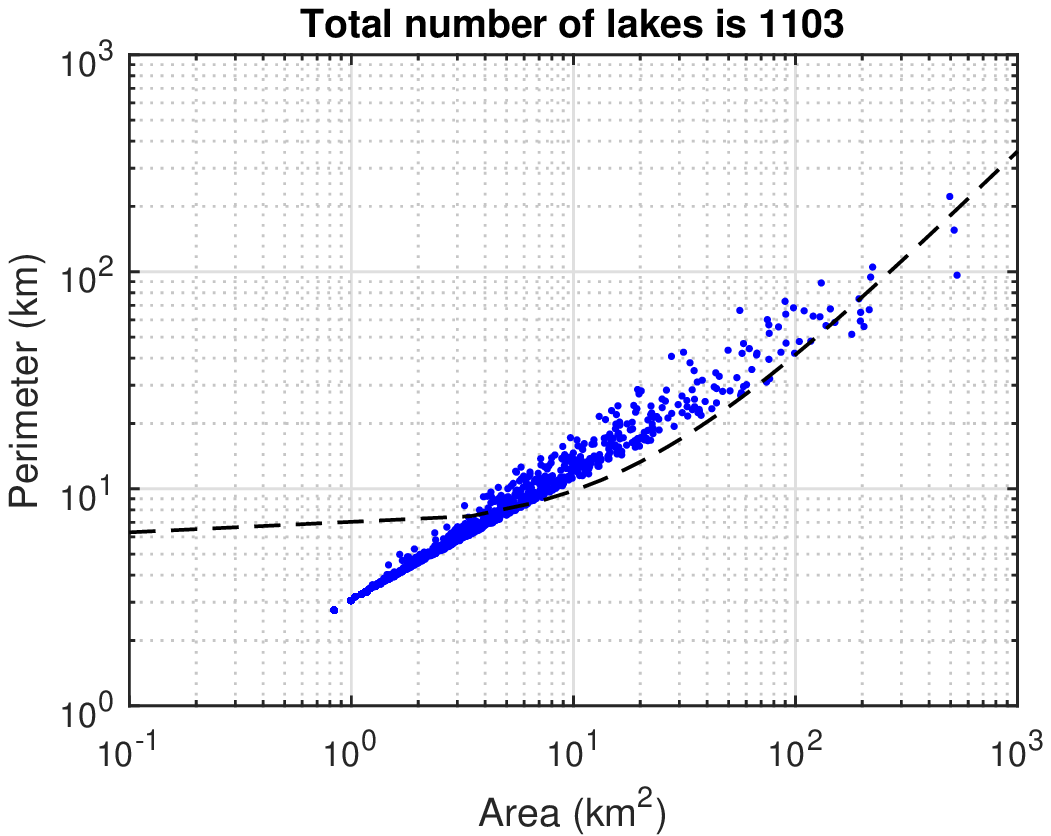}
	\subcaption{} 
\end{minipage}

	\caption{Area--perimeter data plot. The dash line is a linear trend. (\textbf{a}) the plot for the historical topographical map exhibits a change of slope of the liner interpolation curve around a length scale of $30$ km$^2$ in area; (\textbf{b}) the plot for the satellite imagery exhibits a change of slope of the linear interpolation curve around a length scale of $10$ km$^2$ in area.}
	\label{fig-4}      
\end{figure}

There were $864$ tundra lakes detected in the historical topographical map. The area--perimeter values for these lakes were plotted on a log--log scale, and analysis of this plot shows that the linear trend in the data changes slope about $30$ km$^2$ in terms of area (Figure \ref{fig-4}a). These lakes are greater than $0.0008$ km$^2$, which are the smallest lakes that our technique can detect without any false positives. There are numerous features in the map that are smaller than $0.0008$ km$^2$. Visual inspection of our images indicate that these small features are not lakes, but our method could falsely label them as lakes because of their small size. 

There were $1103$ tundra lakes detected in the satellite images. These lakes were greater than $0.8362$ km$^2$. This minimum size was set in order to make the total number of detected lakes from both sources close to each other. This allowed us to compare the fractal dimension of the detected lakes in  Western Siberia using both sources of data. We are able to make the following observations concerning the detection of lakes: (1) if the minimum size of the detected lakes is set to be the same in both the satellite imagery and the historical map, the total number of detected lakes would be much higher in the dataset of satellite images, and this may compromise the comparison of fractal dimension values between the two sources of information; and (2) there are physical limitations on hand-drawn polygons. In the context of the historical map we have analyzed here, these limitations encompass issues ranging from the idea that it is difficult to hand-draw very small objects consistently to the notion that some lake regions may have been colored incorrectly by the person manually rendering the map. We suggest that these factors underlie the observation that the total number of detected lakes is lower in our historical map compared to the satellite imagery in the same region (see Figure \ref{fig-7} below). 

The area–perimeter values for each of these lakes were also plotted on a log–log scale, and analysis of this plot shows that the linear trend in the data changes slope about $10$ km$^2$ in terms of area (Figure~\ref{fig-4}b).




\begin{figure}
	\begin{minipage}{0.495\textwidth}
	\centering
	\includegraphics[width=0.70\linewidth]{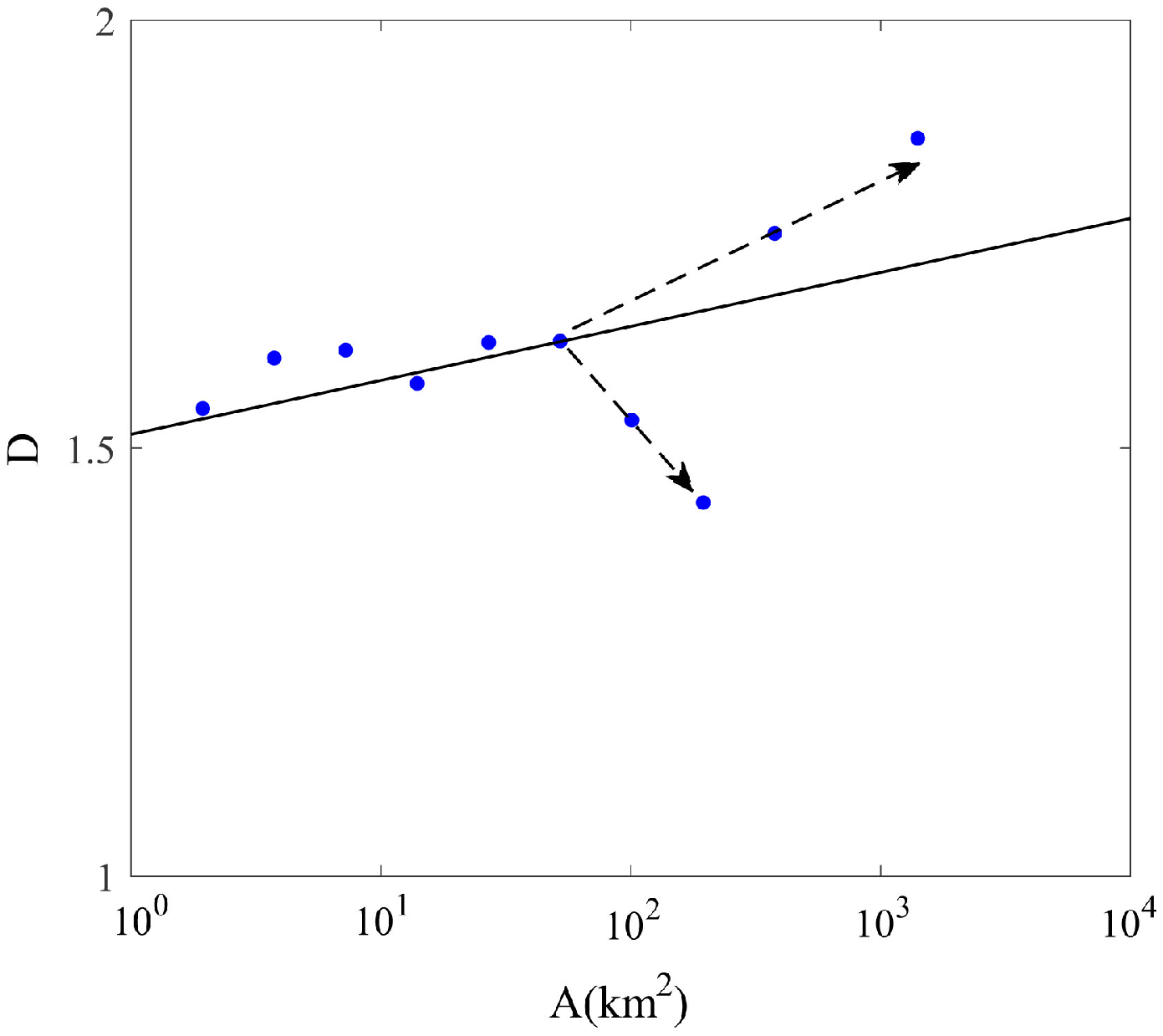}
	\subcaption{} 
	\end{minipage}\hfill
	\begin {minipage}{0.501\textwidth}
	\centering	\includegraphics[width=0.76\linewidth]{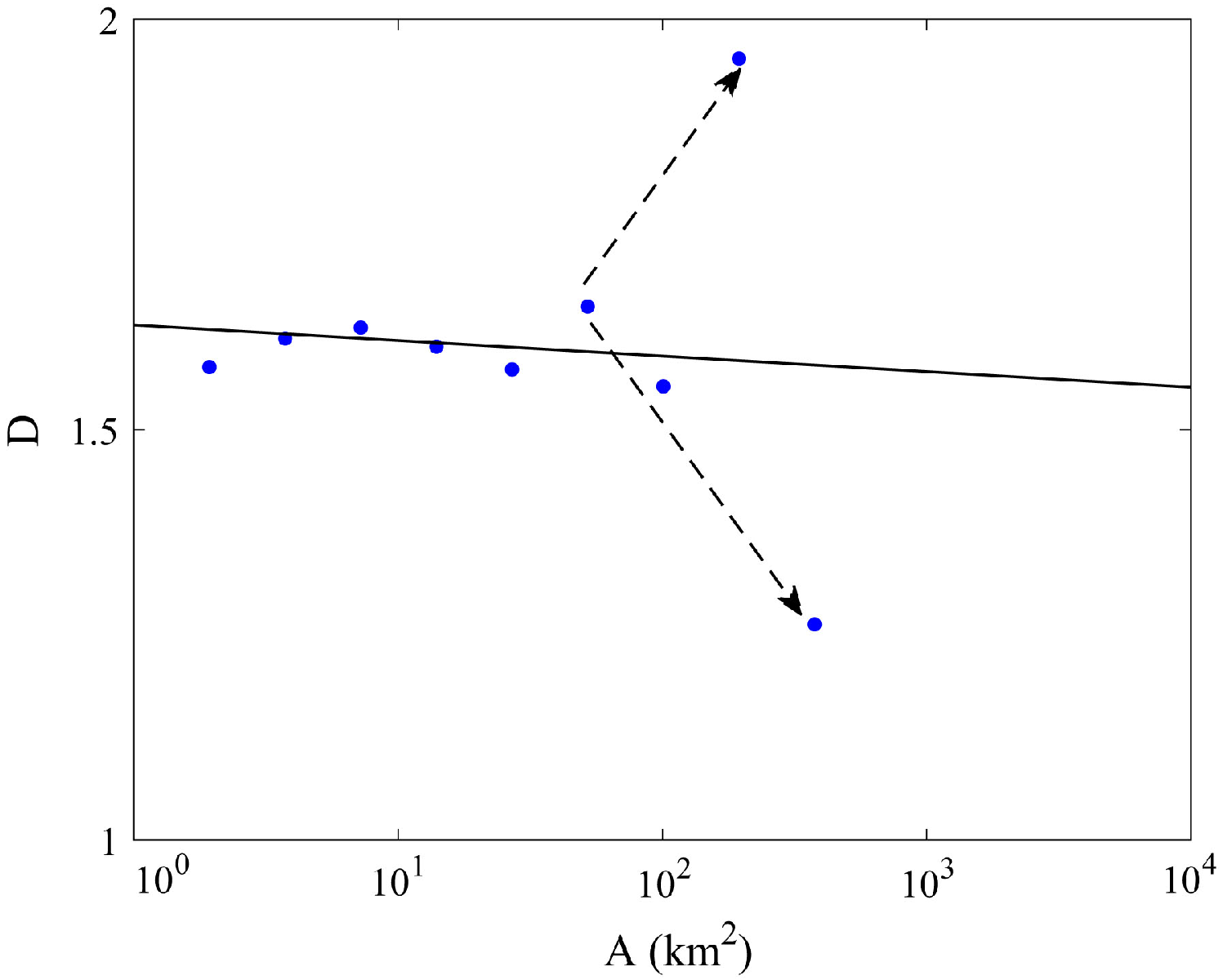}
	\subcaption{}
\end{minipage}

	\caption{Plot of the fractal dimension $D$ as a function of $A$ (log scale). Solid line---a linear interpolation of $D$. Dash line represents a splitting ``bifurcation'' of fractal dimension, which corresponds to more complex geometry of tundra lakes. (\textbf{a}) the plot of the fractal dimension $D$ of $864$ tundra lakes detected on the historical map; (\textbf{b}) the plot of the fractal dimension $D$ of $1103$ tundra lakes detected on the satellite images.}
	\label{fig-5}      
\end{figure}

To investigate this deviation in detail, we computed the fractal dimension $D(A)$ as a function of tundra lake area detected on the historical map. A plot of lake fractal dimension against lake area shows that lakes within the size range 1--70 km$^2$ generally fall on a linear trend-line and have an~average fractal dimension of $1.62$ (Figure \ref{fig-5}a). However, the fractal dimension of lakes larger than $100$ km$^2$ displays interesting behavior: the fractal dimension of some lakes increases above the linear trend line to $1.87$, while the fractal dimension of other lakes falls below the linear trend to $1.43$ (Figure~\ref{fig-5}a). This indicates that the fractal dimension of large tundra lakes in the that historical maps we have analyzed bifurcates.

The fractal dimension of tundra lakes in our satellite images within the size range 1--50 km$^2$ remains constant at about $1.70$ (Figure \ref{fig-5}b). However, the fractal dimension of the lakes larger than $\sim$100 km$^2$ changes dramatically. For some lakes, the fractal dimension is almost $1.95$, whereas others are just $1.31$ (Figure \ref{fig-5}b).




\begin{figure}
	\begin{minipage}{0.51\textwidth}
	\centering
	\includegraphics[width=0.75\linewidth]{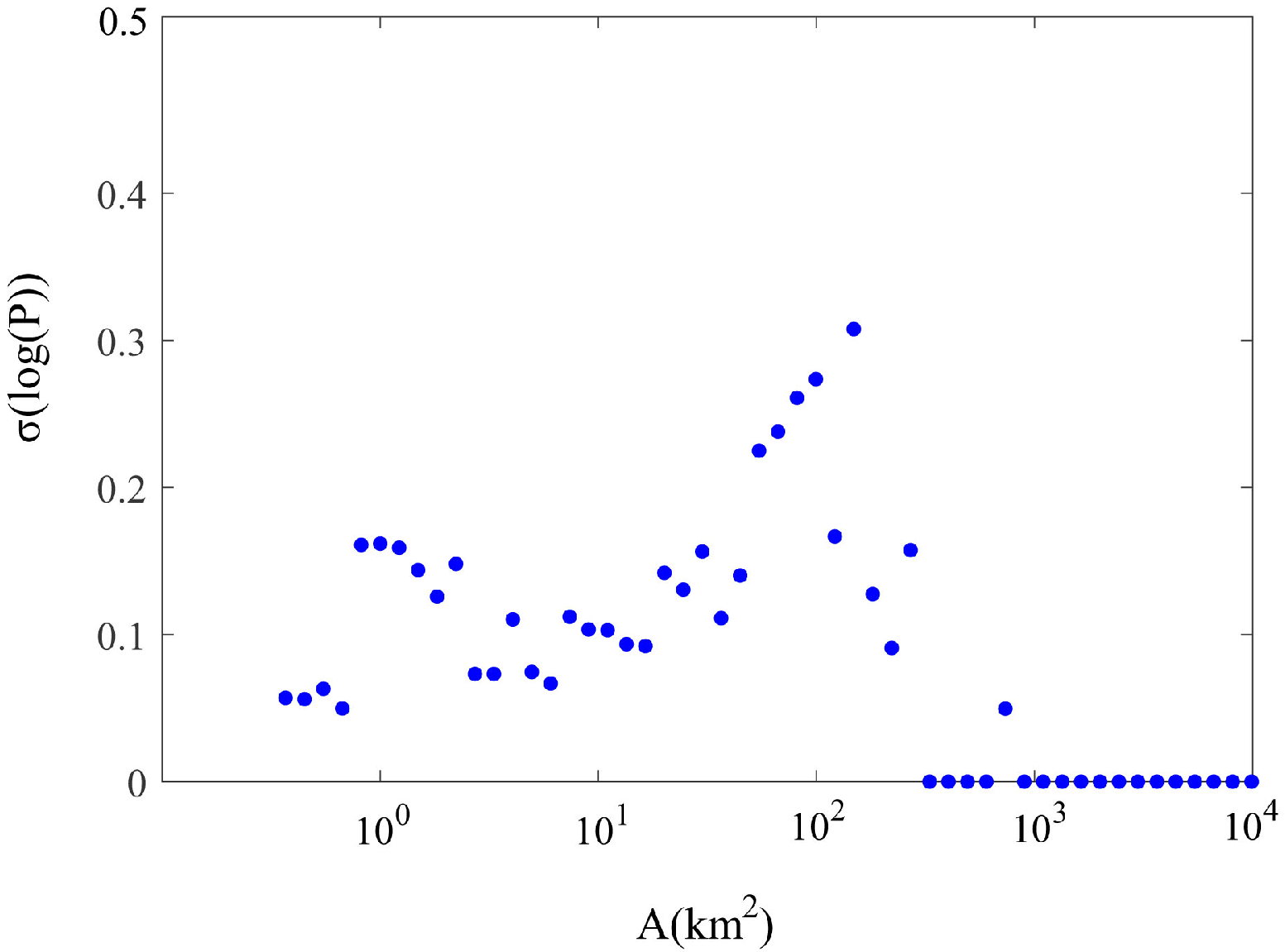}
	\subcaption{} 
	\end{minipage}\hfill
	\begin {minipage}{0.49\textwidth}
	\centering
	\includegraphics[width=0.75\linewidth]{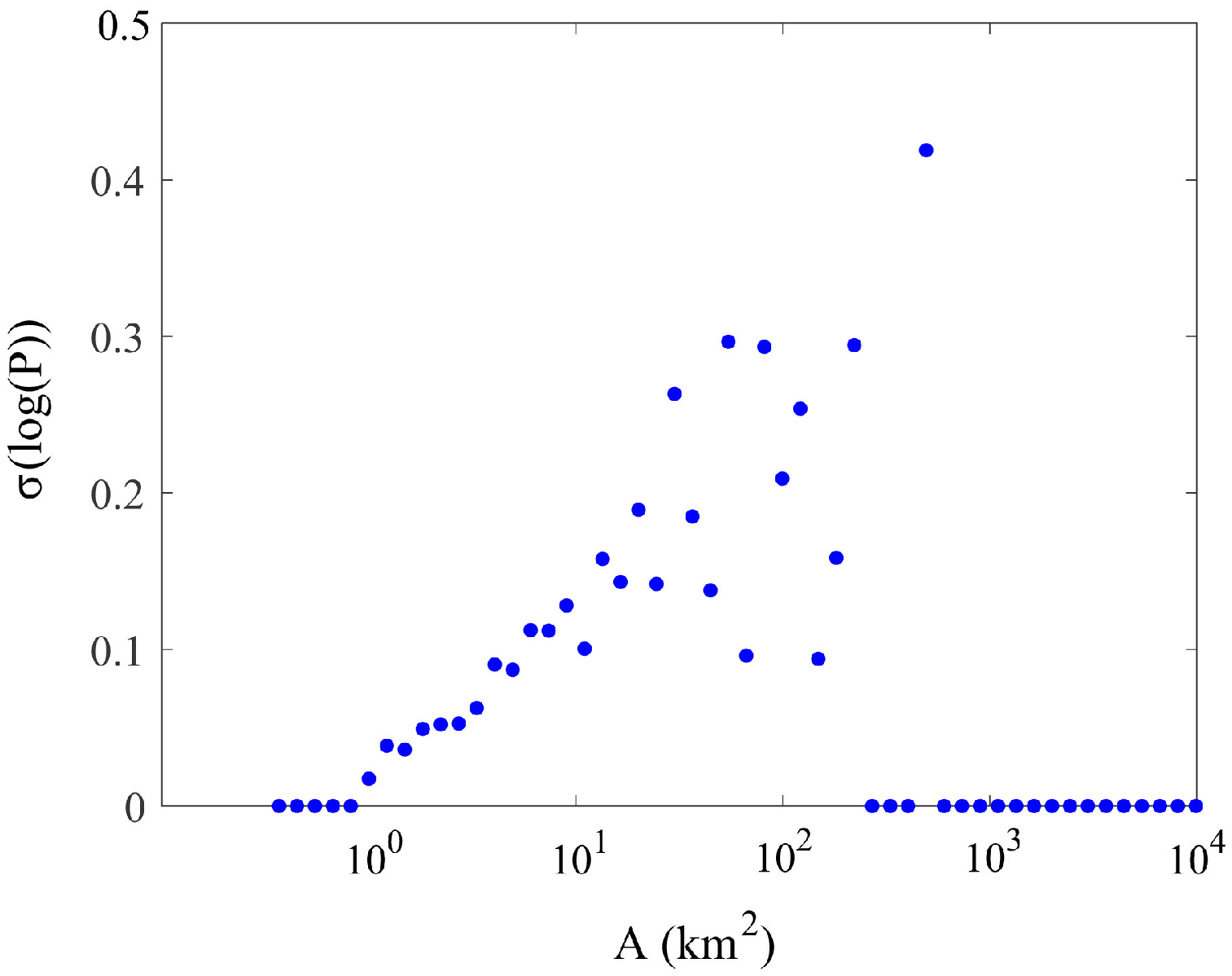}
	\subcaption{} 
\end{minipage}

	\caption{Plot of the variance $\sigma(\log{(P)})$ as a function of $A$ (log scale) for (\textbf{a}) historical map; (\textbf{b}) the satellite imagery.}
	\label{fig-A2}      
\end{figure}

As expected given our method of computation, we obtain few data points for the fractal dimension of the lakes that we have investigated. While this may not look sufficient, especially when we have generated a large data set, these plots ensure that we include only those data points that retain the non-decreasing property of the fractal dimension. To confirm the observed features of lake pattern fractal dimension that we report here, we have employed another parameter called the elasticity and defined the variance $\sigma$ of $\log{(P)}$. Simply stated, variance describes how far a data set is spread out, and~formally this is the average of the squared differences from the mean. The elasticity covers the entire cluster of points on the $(A,P)$-plane. In contrast, the fractal dimension plots show only those data points that maintain the non-decreasing fractal dimension property. The onset of fractal dimension splitting may be identified with the beginning of elasticity cleavage as shown in Figure \ref{fig-A2}. 



\subsection{Statistical Properties of Tundra Lakes Distribution in Historical Maps and Satellite Imagery}

These area--perimeter data allow us to define some statistical properties of lake distribution in order to use this information in future climate modeling.  The distribution of small and large tundra lakes is not stable in time but varies substantially depending on the evolution of climate. What statistical distributions do tundra lakes exhibit depending on historical period? The probability distribution function (PDF) $\textrm{prob}(A)$ in both historical map and satellite imagery can be described by a power law scaling $prob(A)$$\sim$$A^{\zeta}$ with the scaling exponent $\zeta$. 

\begin{figure}
\begin{minipage}{0.5\textwidth}
	\centering
	\includegraphics[width=0.8\linewidth]{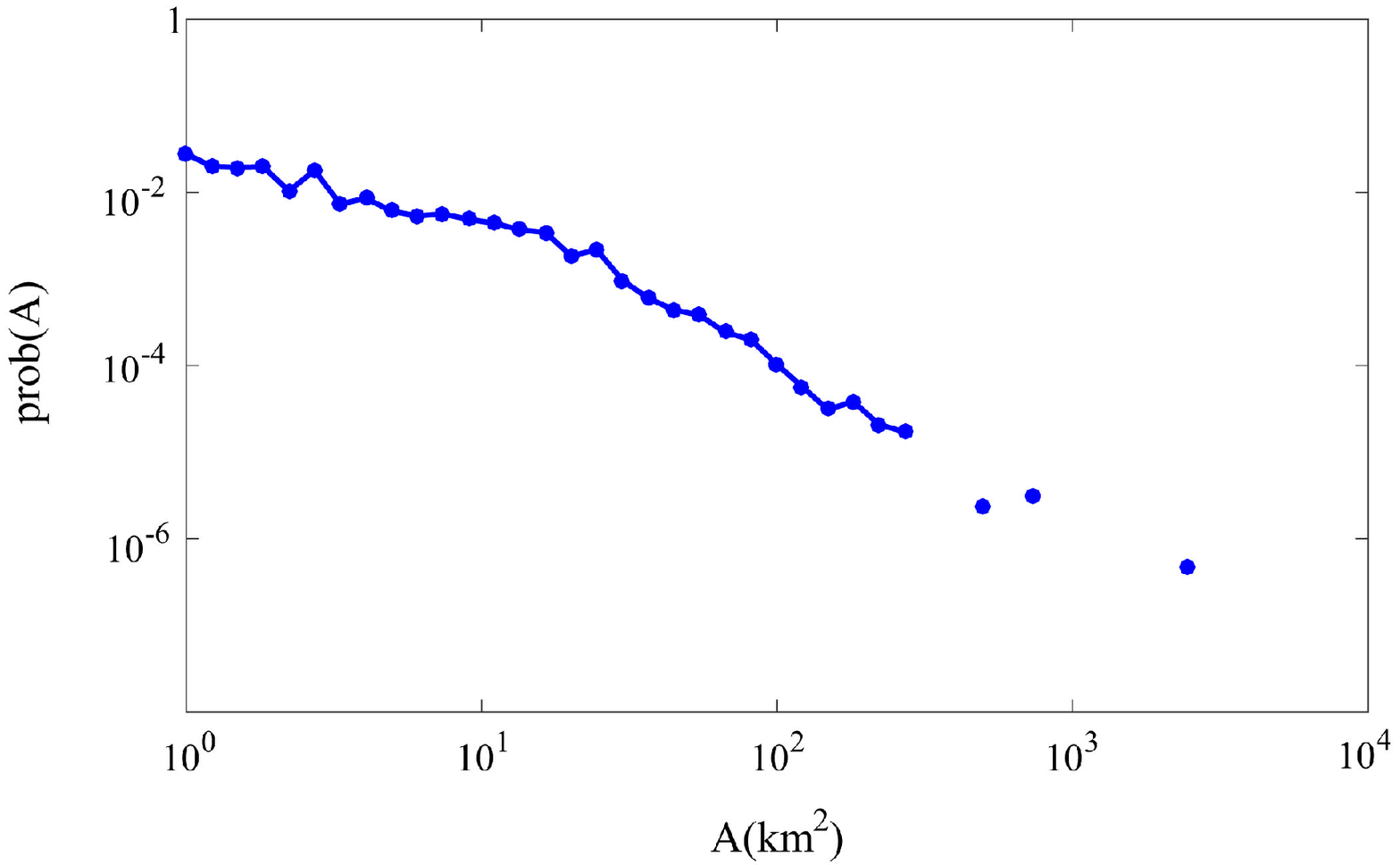}
	\subcaption{} 
	\end{minipage}\hfill
	\begin {minipage}{0.5\textwidth}
	\centering
	\includegraphics[width=0.8\linewidth]{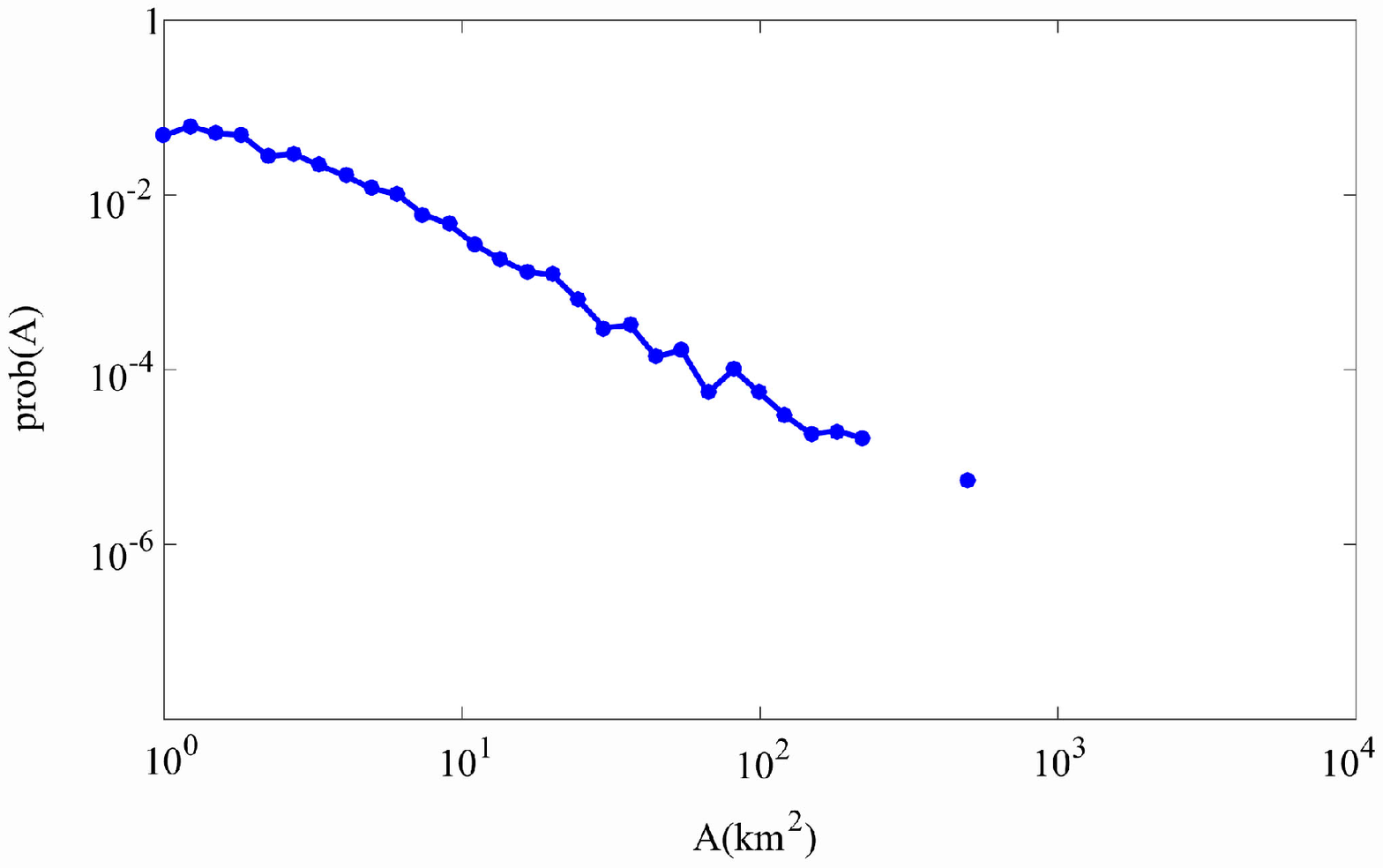}
	\subcaption{} 
\end{minipage}

	\caption{Log--log plot of the lake size distribution function $\textrm{prob}(A)$ for tundra lakes (\textbf{a}) detected from our historical map and (\textbf{b}) detected in satellite imagery derived from the Google Earth Engine.}
	\label{fig-6}      
\end{figure}

	Figure \ref{fig-6}a shows the log--log plot of the PDF with the scaling exponent $\zeta \approx -1.45$  for tundra lakes detected from our historical map. In addition, we can see that the lakes with a length scale larger than $100$~km$^2$ are power-law distributed with a tail exponent ($\tau = 2.28$). Similar exponents have also been observed for other types of lakes on Earth \cite{cae16}. 

Figure \ref{fig-6}b shows the log--log plot of the PDF $\textrm{prob}(A)$  with the scaling exponent $\zeta \approx -1.80$  for tundra lakes detected in satellite imagery derived from the Google Earth Engine. In this case, lakes~with a~length scale larger than $70$ km$^2$ are power-law distributed with a tail exponent ($\tau = 1.93$).

\section{Discussion}
\vspace{-6pt}
\subsection{Image Analysis Effectiveness and Changes in Lake Size over Time}
\begin{figure}
\centering
   \begin{subfigure}[b]{0.7\textwidth}
   \includegraphics[width=1\linewidth]{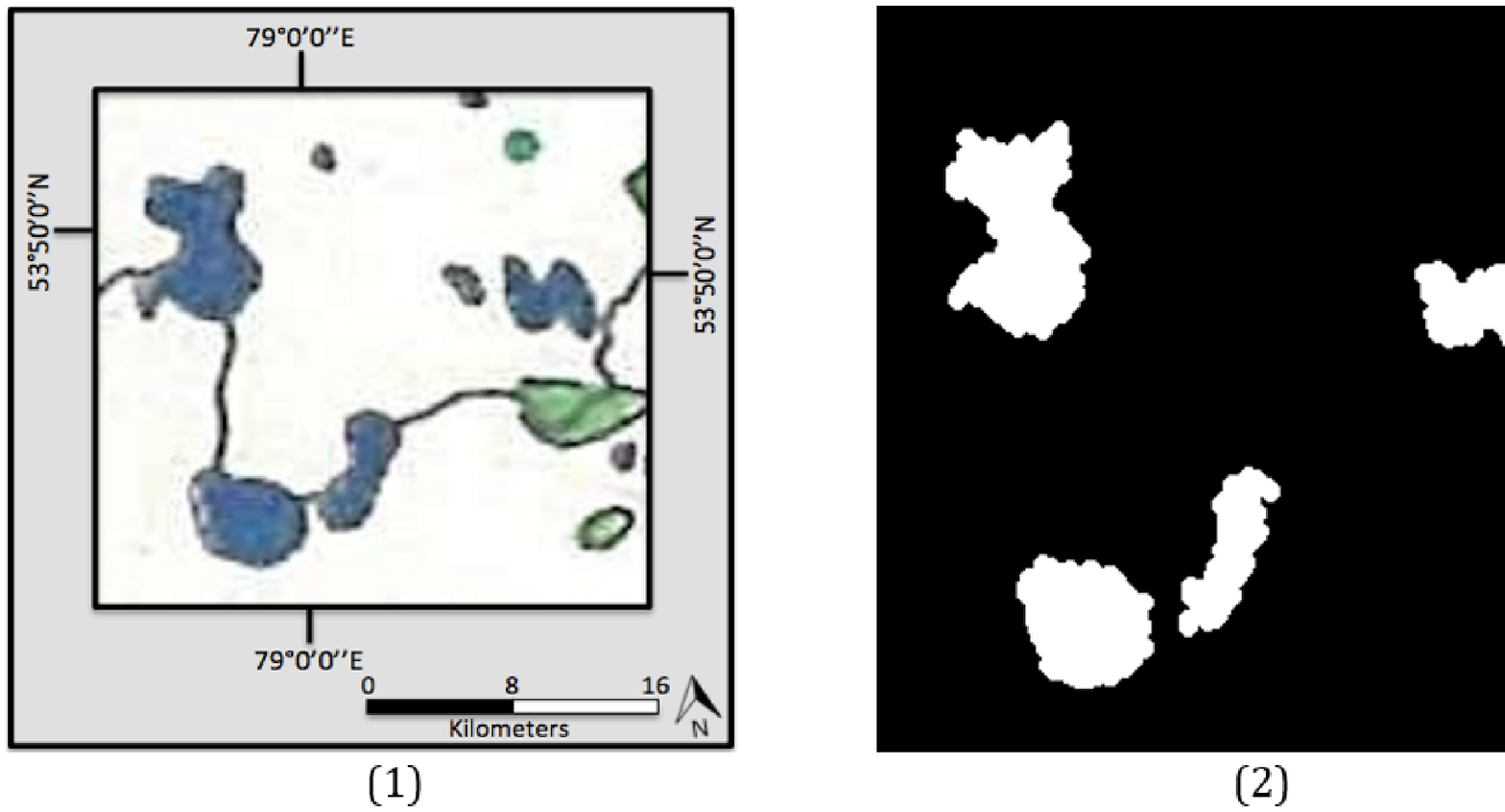}
   \caption{}
\end{subfigure}

\begin{subfigure}[b]{0.7\textwidth}
   \includegraphics[width=1\linewidth]{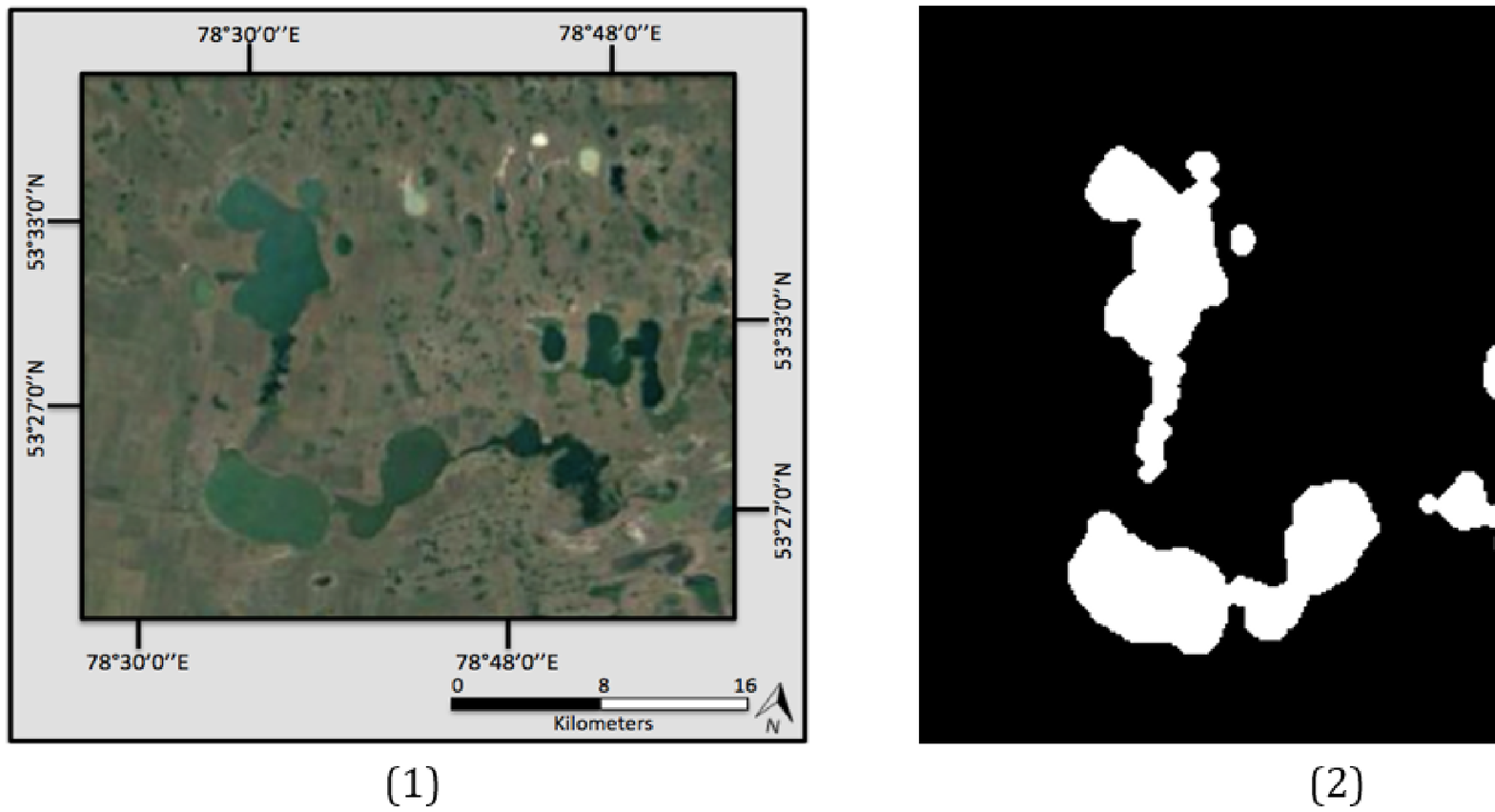}
   \caption{}
\end{subfigure}

\caption{(\textbf{a}) complex lake structure found in the map; (\textbf{b}) complex lake structure detected in a satellite image. (1) input image; (2) segmented image at threshold 0.1; (3) segmented image at threshold 0.05.}
	\label{fig-7}      
\end{figure}

The images of tundra lakes that we have examined in this study, especially the satellite imagery derived from the Google Earth Engine, contain complex information such as color and shape. This~presents a substantial challenge in terms of image segmentation. As mentioned in Section~\ref{sec2.2}, our~image segmentation technique is based on the difference in the intensity values between the pixel of interest and the region of interest. Therefore, small intensity value differences within a lake region itself may cause a single lake to be detected as two or more lakes by our algorithm. This will provide an erroneous number of total detected lakes and will also result in area and perimeter values that are incorrect. Although we attempted to account for this by experimenting with different thresholding values during the image segmentation process, it is possible that these confounding factors mean that some of our detected lake boundaries are artificial. Some of the image segmentation challenges we have faced are shown in Figure \ref{fig-7}.

In this paper, we do not have a time-series of satellite image or a time-series of historical maps with which to investigate how tundra lake geometry has changed through time in our study region. However, we have compared specific lakes that are recorded in both our historical map and our sample of satellite images, and this comparison is located in Table \ref{tab:comparison}. This preliminary analysis illustrates the nature of the variation in lake size in our study area during the time interval that separates the historical map and the satellite imagery. This time interval is around 39 years. These focused example lakes indicate that, in our study region, some lakes have increased in size over time, whereas others have decreased in size over time, and changes in lake size during this time interval can be up to half the size of the lake as recorded in the historical map. 

Clearly, it is difficult to draw substantial conclusions from a comparison of this nature because the two data sources are so different, each with its own advantages and limitations. For instance, Figure~\ref{fig-7} suggests that there are some lakes in the Google Earth Engine imagery (see Figure \ref{fig-7}b) that do not exist in the historical map (see Figure \ref{fig-7}a). However, visual comparison of the two datasets indicates that the lakes were in fact present in the historical map, but they were mislabeled. Specifically, they~were colored green instead of dark blue, which is the color used for the vast majority of lakes in this map, and so they were missed by our algorithm during the segmentation of the historical map. We have included this comparison to highlight an emerging challenge in this research domain: the~integration and comparison of different sources of information on tundra lake geometry. This~challenge arises because certain types of data on historical lake geometry, such as the historical map used here, may~simply not exist for the present day. Similarly, satellite data for a given region may span a decade or so, but other sources of information, such as maps or aerial photographs, could provide valuable historical context for such data.  

\vspace{-12pt}
\begin{table}
\centering
\caption{Lake area and perimeter comparison.}
\label{tab:comparison}
\begin{tabular}{p{1.5cm}<{\centering}p{1.7cm}<{\centering}p{1.1cm}<{\centering}p{1.5cm}<{\centering}p{1.7cm}<{\centering}p{1.5cm}<{\centering}p{1.1cm}<{\centering}p{1.5cm}<{\centering}}
\toprule

\multicolumn{4}{c}{\textbf{Historical Map}}  & \multicolumn{4}{c}{\textbf{Google Earth Engine}}\\
\midrule

 \textbf{Original Image} & \textbf{Binary Image} &  \textbf{Area (km}\boldmath$^2$\textbf{)} &  \textbf{Perimeter (km)} &  \textbf{Original Image} & \textbf{Binary Image} &  \textbf{Area (km}\boldmath$^2$\textbf{)} &  \textbf{Perimeter (km)} \\
\midrule

\vspace{.1mm} \includegraphics[scale=0.37]{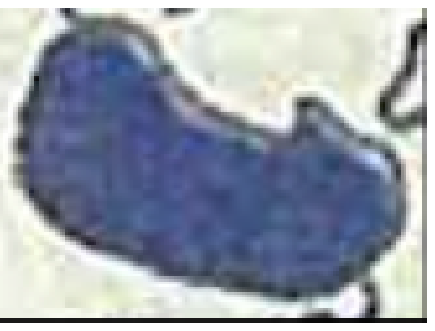} & \vspace{.1mm} \includegraphics[scale=0.4]{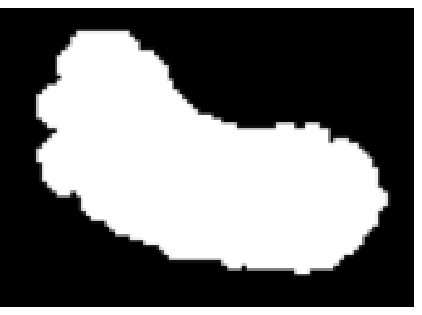} & 
\vspace{4mm}\small{116.883}  & \vspace{4mm}\small{48.309} & 
\vspace{.01mm} \includegraphics[scale=0.35]{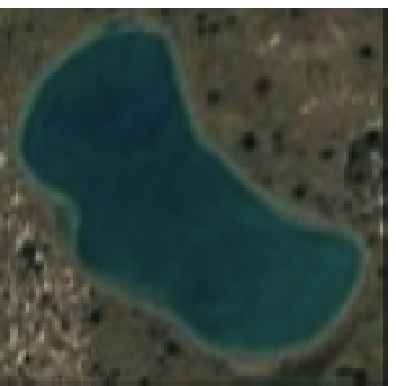} & \vspace{.01mm} \includegraphics[scale=0.28]{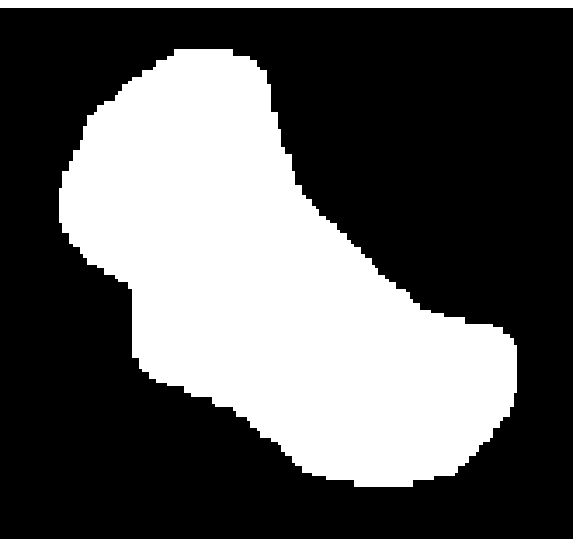} & 
\vspace{4mm}\small{98.920} & \vspace{4mm}\small{41.899} \\ [10pt]
\midrule

\vspace{.1mm}\includegraphics[scale=0.4]{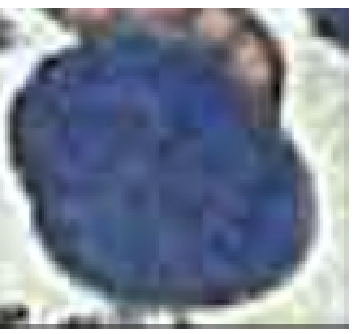} & \vspace{.1mm}\includegraphics[scale=0.37]{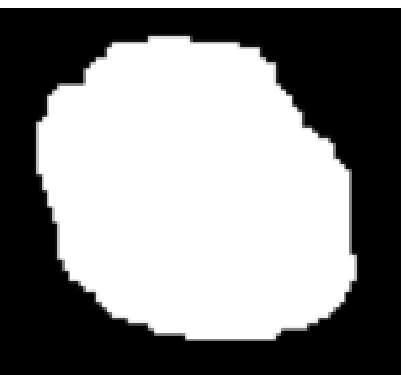} & 
 \vspace{5mm}\small{95.916} & \vspace{5mm} \small{35.515} & \vspace{.1mm}\includegraphics[scale=0.41]{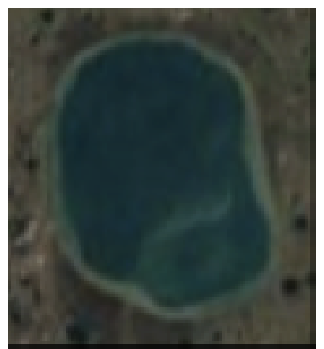} & \vspace{.1mm}\includegraphics[scale=0.35]{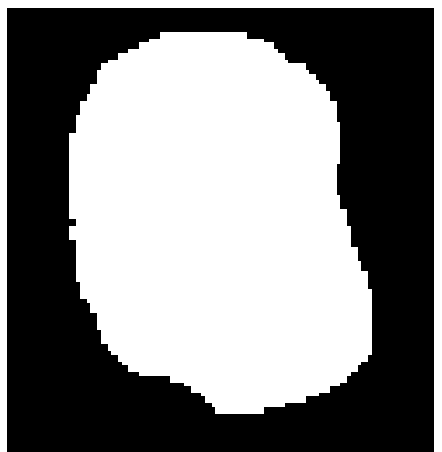} & \vspace{5mm}\small{76.275} & \vspace{5mm}\small{32.057} \\ [10pt]
\midrule

\vspace{.1mm}\includegraphics[scale=0.45]{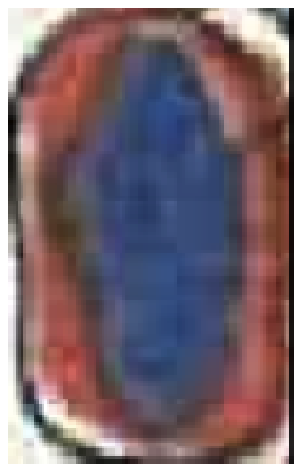} & \vspace{.1mm}\includegraphics[scale=0.5]{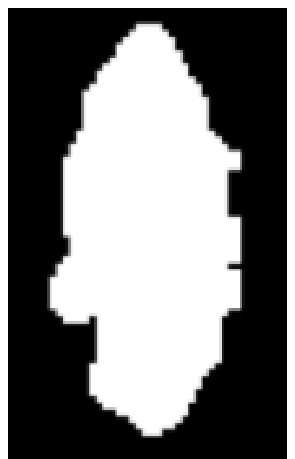} & \vspace{9mm}\small{52.416} & \vspace{9mm}\small{30.615} & \vspace{.1mm}\includegraphics[scale=0.4]{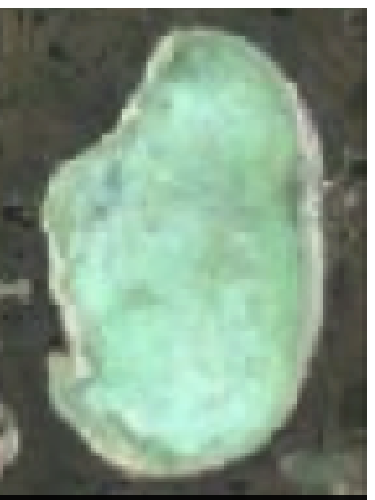} & \vspace{.1mm}\includegraphics[scale=0.32]{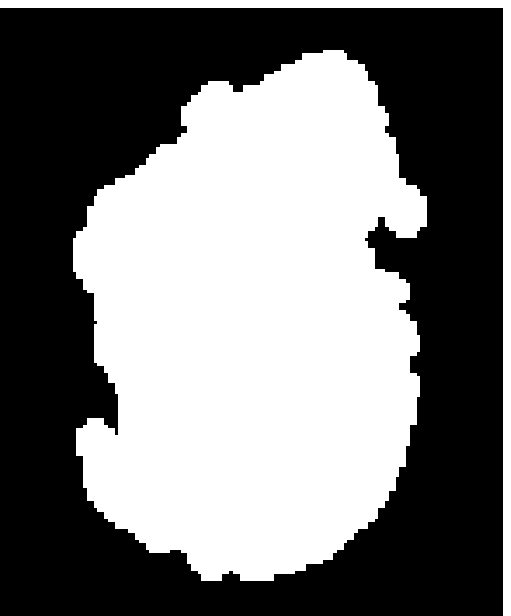} & \vspace{9mm}\small{117.316} & \vspace{9mm}\small{47.914}\\ [10pt] 
\midrule

\vspace{.1mm}\includegraphics[scale=0.36]{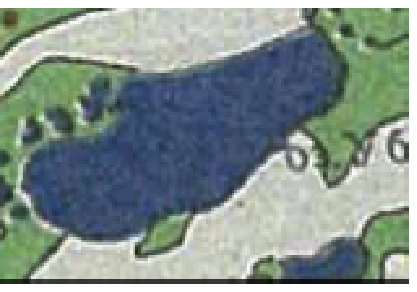} & \vspace{.1mm}\includegraphics[scale=0.38]{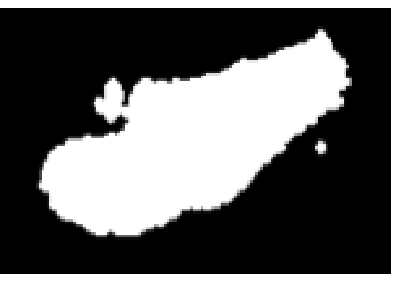} & \vspace{3mm}\small{478.553} & \vspace{3mm}\small{115.059} & \vspace{.1mm}\includegraphics[scale=0.37]{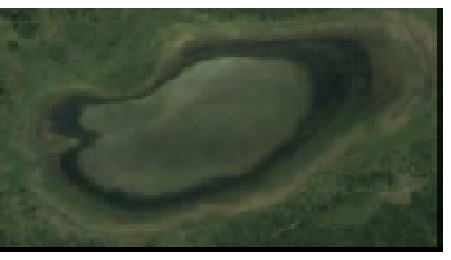} & \vspace{.1mm}\includegraphics[scale=0.2]{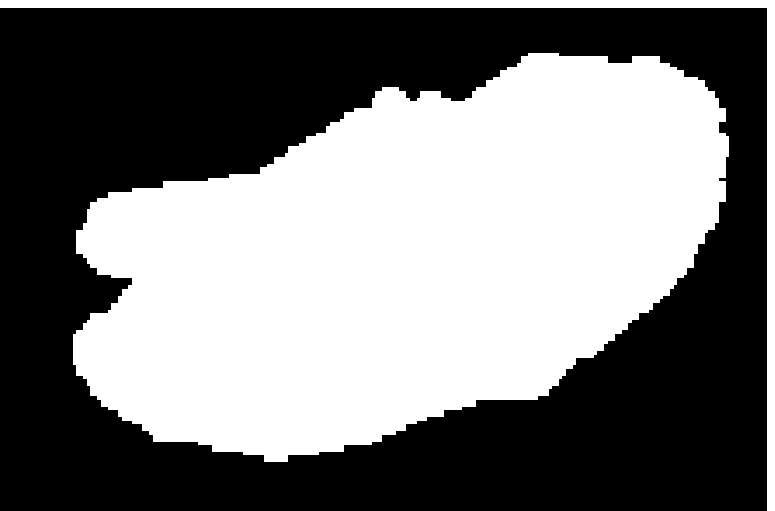} & \vspace{3mm}\small{159.262} & \vspace{3mm}\small{53.789}\\[10pt]
\bottomrule

\end{tabular}
\end{table}

\subsection{Phase Transitions and the Complex Geometry of Tundra Lake Patterns}
The fractal dimension of the tundra lake patterns in both the historical map and satellite images indicate that lake patterns with a short length scale ($100$ km$^2$) have a non-changing fractal dimension of around 1.6–1.7. This is in agreement with previous research on the fractal dimension of some elements of the Siberian tundra landscape (where the fractal dimension of tundra lakes is generally 1.68) \cite{bak07}, and is typical of Earth's lakes in general, which have no tendency to change fractal dimension over time and also remain close to a limit \cite{cae16}.

Although several natural patterns such as clouds and rivers exhibit constant fractal dimension, there is a system of melt ponds on Arctic sea ice that is characterized by dramatic geometrical variability. Melt ponds on Arctic sea ice develop due to phase transitions (sea-ice to ocean-water) as a result of increased temperature and radiative forcing. It has been shown that the fractal dimension of Arctic melt ponds shifts abruptly from $1$ to $2$ as ponds grow in size, with the critical transition zone around $100$ m$^2$ \cite{hohenegger13}. Fundamentally, the thawing of permafrost and development of tundra lakes is also a~phase transition phenomenon, where a solid turns to a liquid, albeit on a large regional scale and over a~period of time that depends on environmental forcing and other more local factors. We~suggest that the "bifurcation" we observe in the fractal dimension of the tundra lakes that we have studied represents a~transition phenomenon that is analogous to that seen in melt ponds \cite{hohenegger13}. 

However, it should be emphasized that, in the case of the tundra lakes we have studied, the fractal dimension ``bifurcates'', whereas, in the case of Arctic melt ponds, the fractal dimension only increases in a nonlinear fashion \cite{hohenegger13}. This may imply that the geometrical evolution of tundra lakes is more complex than Arctic melt ponds. We suggest that the bifurcation of fractal dimension in the tundra lakes we have investigated may be explained by the drainage of large lakes to the wider regional hydrological network. In particular, larger lakes become fragmented following partial drainage \cite{and15} and this leads to a decline in lake area, and thus to a lake that itself has a lower fractal dimension. This~process is characteristic of a lake transforming to peatland \cite{kir08}. 

Tundra lakes with a fractal dimension higher than average and close to 2 have a tendency to be self-similar with respect to their area--perimeter relationships. We are able to recognize a portion of a~lake that is connected to a complex lake system, such that the perimeter to area ratio of the entire lake is approximately the same as that of the portion. For example, by computing the perimeter to area ratios
for a whole lake and a portion of that lake (Figure \ref{fig-7}b), we find $P/A\approx 1.2$ km$^{-1}$ for both the lake and its portion. The measurement of complexity properties such as this may allow tundra lakes to be classified on the basis of their structure and fractal dimension.

Our statistical analyses of the size-frequency distribution of tundra lakes on the Arctic land surface also indicate that these lakes follow a power law. This highlights that a first-order property of the tundra landscape is the high abundance of very small lakes together with considerably smaller numbers of larger lakes. Such knowledge may be used in stochastic models of methane emissions from boreal permafrost \cite{huis11,sudakov15}. For example, in \cite{sudakov15}, the authors assume that the lake size distribution follows the Pareto distribution (a power law with specific exponents). In such cases, the scaling exponents of the power law could be found definitely from observations using the image analyses of tundra lakes that we have developed here.

Finally, our results indicate that images derived from the free access platform Google Earth Engine can be used to detail and diagnose the fractal dimension of tundra lakes in different regions of the Arctic. By virtue of its reliance on an open-access source of imagery, our approach may represent a~convenient blueprint of large-scale surveys of tundra lake geometry in other regions of the world. While multi-spectral Landsat data are widely used  for analysis of high resolution tundra lake dynamics~\cite{nitze17}, we suggest that our approach of using open access imagery is also appropriate to investigate spatiotemporal dynamics of tundra lakes.  At the very least, it provides a means to track lake size change through time that will be useful for numerical climate model parametrization \cite{stepanenkoa11,tan15}.



\section{Conclusions}
Our study of tundra lake geometry is motivated by an overarching desire to understand the spatiotemporal dynamics of tundra lake patterns under climatic change. This is important because tundra lakes represent a significant source of methane to the atmosphere, and they change shape and increase in size as permafrost thaws. In this methodological paper, we have taken a step towards this goal by analyzing the geometrical properties of tundra lakes in the Russian High Arctic. These~lakes were recorded by historical maps and by satellite imagery derived from the Google Earth Engine. Our~specific conclusions are as follows:
\begin{enumerate}[leftmargin=*,labelsep=4.9mm]
\item	[(1)] An image processing strategy that combines color thresholding and region growing allows us to segment historical maps and satellite images. This isolates lakes from other information in these images. In the case of historical maps, this other information includes the distribution of wetlands, and in the case of satellite images includes vegetation, soil and other components of the land surface.
\item	[(2)] The fractal dimension of lakes within the size range 1--70 km$^2$ observed in historical maps is on average $1.62$ (Figure \ref{fig-5}a). However, the fractal dimension of lakes larger than $100$ km$^2$ displays bifurcating behavior: the fractal dimension of some of these large lakes is $1.87$ and the fractal dimension of some of these large lakes is $1.43$ (Figure \ref{fig-5}a). 
\item	[(3)] Similar behavior in the fractal dimension of lakes was observed in satellite imagery. The fractal dimension of lakes sized 1--50 km$^2$ is around $1.70$ (Figure \ref{fig-5}b), but the fractal dimension of the lakes larger than $\sim$100~km$^2$ ranges from $1.31$ to $1.95$ (Figure \ref{fig-5}b).
\item	[(4)] Area--perimeter measurements for each of the individual lakes in our analyses show that, for lakes in our historical maps, those with a length scale larger than $100$ km$^2$ are power-law distributed with a tail exponent ($\tau = 2.28$). For lakes in the satellite images we have analyzed, those with a~length scale larger than $70$ km$^2$ are also power-law distributed with a tail exponent ($\tau = 1.93$). Similar exponents have been observed for other lakes on the Earth surface \cite{cae16}. 
\item    [(5)] We have undertaken a preliminary analysis of the changes in lake size that have taken place in the ~39 year interval between the production of the historical map we have used and the satellite images we have examined. We find that some lakes in our study region have increased in size over time, whereas others have decreased in size over time. Changes in lake size during this time interval can be up to half the size of the lake as recorded in the historical map. 
\end{enumerate}


\vspace{6pt}
\section*{Acknowledgments}
A.E. and M.G. gratefully acknowledge support from the graduate student cost sharing funding scheme realized by the Office of the Dean of the College of Arts and Sciences, the Office of the Dean of School of Engineering, the Department of Physics, the Department of Electrical and Computer Engineering and Graduate Academic Affairs, University of Dayton. We also thank the National Science Foundation (NSF) Math Climate Research Network for their support of this work. We acknowledge support from the the Russian Foundation for Basic Research (RFBR) under the Grant \#16-31-60070 mol\_a\_dk. Finally, we gratefully acknowledge support from the Division of Mathematical Sciences at the U.S. NSF through Grant DMS-0940363. 
We would like to thank Vijayan Asari (University of Dayton, Dayton, OH, USA) for detailed discussion of the proposed method of image analysis. In addition, we thank Yi-Ping Ma (Northumbria University, Newcastle, UK) for useful discussion of fractal dimension computation.

\section*{Author Contributions} I.S. and L.M. proposed the idea, designed the framework of research and developed the data analysis. A.E. proposed the method of image analysis. A.E., M.G. and I.S conducted the data analysis. T.K.~made the materials review. All authors contributed significantly to writing the manuscript.

\section*{Conflicts of Interest} The authors declare no conflict of interest. The founding sponsors had no role in the design of the study; in the collection, analyses or interpretation of data; in the writing of the manuscript, and in the decision to publish the results.











\begin{thebibliography}{31}
\bibitem{duarte12}
Duarte, C.M.; ~Lenton, T.M.; ~Wadhams, P.; ~Wassmann, P. ~Abrupt climate change in the Arctic.
\emph{Nat.~Clim.~Change}  ~{\bf 2012}, \emph{2}, 60--62.

\bibitem{hinzman13}
Hinzman, L.D.; ~Deal, C.J.; ~McGuire, A.D.; ~Mernild, S.H.; ~Polyakov, I.V; ~Walsh, J.E. ~Trajectory of the Arctic as an integrated system. 
\emph{~Ecol. Appl.} {\bf 2013}, \emph{23}, 1837--1868.

\bibitem{lenton08}
Lenton, T.M.; ~Held, H.; ~Kriegler, E.; ~Hall, J.W.; ~Lucht, W.; Rahmstorf, S.; Schellnhuber, H.J. ~Tipping elements in the Earth’s climate system. 
\emph{~Proc. Natl. Acad. Sci. USA} ~{\bf 2008}, \emph{105}, 1786--1793.

\bibitem{schu15}
Schuur, E.A.G.; ~McGuire, A.D.; ~Schadel, C.; ~Grosse, G.; ~ Harden, J.W.; ~Hayes, D.J.; ~Hugelius, G.; Koven,~C.D.; ~Kuhry, P.; ~Lawrence, D.M.; et al. ~Climate change and the permafrost carbon feedback. \emph{Nature}~{\bf 2015}, \emph{520}, 171--179.

\bibitem{hope}
Hope, C.; Schaefer, K. Economic impacts of carbon dioxide and methane released from thawing permafrost. \emph{~Nat. Clim. Change} ~{\bf 2016}, \emph{6}, 56--59, doi:10.1038/nclimate2807.
\bibitem{Wik16}
Wik, M.; ~Ruth, K.V.; ~Anthoney, K.W.; ~MacIntyre, S.; ~Bastviken, D. ~Climate-sensitive northern lakes and ponds are critical components of methane release. \emph{~Nat. Geosci.}~{\bf 2016}, \emph{9}, 99--105.

\bibitem{walt06}
Walter, K.M.; ~Zimov, S.A.; ~Chanton, J.P.; ~Verbyla, D.; ~Chapin, F.S., III. ~Methane bubbling from Siberian thaw lakes as positive feedback to climate warming. 
\emph{Nature} ~{\bf 2006}, \emph{443}, 71--75.

\bibitem{golubyatnikov13}
Golubyatnikov, L.L.; ~Kazantsev, V.S. ~Contribution of tundra lakes in Western Siberia to the atmospheric methane budget. \emph{{Izv.} Atmos. Ocean. Phys.} ~{\bf 2013}, \emph{49},  395--403.

\bibitem{and15}
Andresen, C.G.; ~Lougheed, V.L. ~Disappearing Arctic tundra ponds: Fine-scale analysis of surface hydrology in drained thaw lake basins over a 65 year period (1948--2013). 
\emph{J. Geophys. Res. Biogeosci.} {\bf 2015}, \emph{120}, 1--14.

\bibitem{leg13}
Legleiter, C.J.; Tedesco, M.; Smith, L.C.; Arp, C.D.; Behar, A.E.; Overstreet, B.T. Mapping the bathymetry of supraglacial lakes and streams on the Greenland ice sheet using field measurements and high-resolution satellite images. \emph{Cryosphere} {\bf 2014}, \emph{8}, 215--228.
\bibitem{hohenegger13}
Hohenegger, C.; ~Alali, B.;  ~Steffen, K.R.; ~Perovich,  D.K.; ~Golden, K.M. ~Transition in the fractal geometry of Arctic melt ponds.\emph{~Cryosphere} {\bf 2013}, \emph{6}, 1157--1162.

\bibitem {kar13}
Karlsson, J.M.; ~Lyon, S.W.; ~Destouni, G.~Temporal behavior of lake size-distribution in a thawing permafrost landscape in northwestern Siberia. 
\emph{Remote Sens.} {\bf 2014}, \emph{6}, 621--636.

\bibitem{polishchuk15}
Polishchuk, Y.M.; ~Bryksina, N.A.; Polishchuk, V.Y. ~Remote analysis of changes in the number of small thermokarst lakes and their distribution with respect to their sizes in the cryolithozone of Western Siberia. 
\emph{{Izv}. Atmos. Ocean. Phys.}~{\bf 2015}, \emph{51}, ~999--1006.

\bibitem{nitze17}
Nitze, I.; ~Grosse, G.; ~Jones, B.M.; ~Arp, C.D.; ~Ulrich, M.; ~Fedorov, A.; ~Veremeeva, A. ~A  Landsat based trend analysis of lake dynamics across northern permafrost regions. 
\emph{Remote Sens.}~{\bf 2017}, \emph{9},~640--668.

\bibitem{man17}
Mander, L.; ~Dekker, S.C.; ~Li, M.; ~Mio, W.; ~Punyasena, S.W.; ~Lenton, T.M. ~A morphometric analysis of vegetation patterns in dryland ecosystems. 
\emph{R. Soc. Open Sci.} {\bf 2017}, \emph{4}, doi:10.1098/rsos.160443.
\bibitem{kim13}
Kim, D.J.; Jung, H.S. Melt Pond mapping with high-resolution SAR: The first view. \emph {Proc. IEEE.} {\bf 2013}, \emph{101}, 748--758.

\bibitem{polishchuk13}
Polishchuk, Y.M.; ~Bogdanov, A.N.; ~Polishchuk, V.Y.; ~Manasypov, R.M.; ~Shirokova, L.S.; Kirpotin, S.N.; Pokrovsky, O.S.~Size distribution, surface coverage, water, carbon, and metal storage of thermokarst lakes in the permafrost zone of the Western Siberia lowland.
\emph{Water} {\bf 2017}, \emph{9}, 228.

\bibitem {kar14}
Karlsson, J.M.; ~Lyon, S.W.; ~Destouni, G. ~Temporal behavior of lake size-distribution in a thawing permafrost landscape in northwestern Siberia. 
\emph{Remote Sens.} ~{\bf 2014}, \emph{6}, 621–636. 

\bibitem{h-map}
Romanova, E.A. \emph{Tipologicheskaya Karta Bolot Zapadno-sibirskoi Ravniny. [Typological Map of the Wetlands of the West Siberian Plain]};  The State Hydrological Institute: Leningrad, Russia, 1977. (In Russian) 

\bibitem{iv76}
Ivanov, K.E.; Novikov, S.M.~\emph{Bolota Zapadnoi Sibiri, Ikh Stroenie I Gidrologicheskiy Rezhim. [Bogs of West Siberia, their Structure and Hydrology]};~Gidrometeoizdat: Leningrad, Russia, 1976. (In Russian)
\bibitem{rom70}
Romanova, E.A. {Deshifrirovaniye aerofotosnimkov i sostavleniye krupnomasshtabnykh tipologicheskikh kart bolot Zapadnoy Sibiri. [Decoding of aerial photographs and creation of large-scale typological maps of bogs in the Western Siberia]}.  In~\emph{Krupnomasshtabnoye Kartografirovaniye Rastitel'nosti. [Large-Scale Vegetation Mapping]};~Nauka: Novosibirsk, Russia, 1970; pp. 118--123. (In Russian)


\bibitem{g-map}
The Imagery of the Western Siberia ($70^{\circ}24^{\prime}27.16^{\prime\prime}$ N and $71^{\circ}53^{\prime}11.70^{\prime\prime}$ E). Google Earth Engine. Issued 30 December 2016. 
(accessed on August 2017).

\bibitem{ge}
Gorelick, N.; Hancher, M.; Dixon, M.; Ilyushchenko, S.; Thau, D.; Moore, R. Remote sensing of environment Google Earth Engine: Planetary-scale geospatial analysis for everyone. \emph{Remote Sens. Environ.}~{\bf 2017}, doi:10.1016/j.rse.2017.06.031. 

\bibitem{grl}
Deines, J.M.; Kendall, A.D.; Hyndman, D.W.  Annual Irrigation Dynamics in the U.S. Northern High Plains Derived from Landsat Satellite Data. \emph{Geophys. Res. Lett.}~{\bf 2017}, doi:10.1002/2017GL074071 

\bibitem{svm}
Boser, B.E.; ~Guyon, I.M.; ~Vapnik, V.N. ~A training algorithm for optimal margin classifiers. In Proceedings of the Fifth Annual ACM Workshop on Computational Learning Theory, Pittsburgh, PA, USA, 27--29 July 1992. 
\bibitem{17}
Gonzalez, R.C.;  Woods, R.E.  \emph{Digital Image Processing}; Prentice Hall: Upper Saddle River, NJ, USA, 2002. 

\bibitem{regiongrowingcode}
Kroon, D. Region Growing Code. Available online: \url {https://www.mathworks.com/matlabcentral/fileexchange/19084-region-growing} (accessed on 1 June 2017).

\bibitem{18}
Dillencourt, M.B.; ~Samet, H.; ~Tamminen, M.~A general approach to connected-component labeling for arbitrary image representations.
\emph{J. ACM} {\bf 1992}, \emph{39}, 253--280. 

\bibitem{mandelbrot77}
Mandelbrot, B.B. \emph{Fractals: Form, Chance and Dimension}; ~W.H. Freeman and Company: San Francisco, CA, USA, 1977.

\bibitem{cae16}
Cael, B.B; ~Seekell, D.A. The size-distribution of Earth’s lakes. {\em Sci. Rep.}~{\bf 2016}, {\em 6}, doi:10.1038/srep29633.
\bibitem{bak07}
Balkhanov, V.K.; ~Lukhneva, O.F.; ~Kusner, Y.S.; Bashkuyev, Y.B.~Fractal dimensionality of the Lena river delta and tundra lakes in Yakutia. {\em Geogr. Nat. Resour.} {\bf 2008}, {\em 2},~153--157.


\bibitem {kir08}
Kirpotin, S.;~Polishchuk, Y.; ~Zakharova, E.; ~Shirokova, L.; ~Pokrovsky, O.; ~Kolmakova, M.; ~Dupre, B. ~One of possible mechanisms of thermokarst lakes drainage in West-Siberian North. \emph{Int. J. Environ. Stud.}~{\bf 2008}, \emph{65}, 631--635.

\bibitem{huis11}
Van Huissteden, J; ~Berrittella, C.; ~Parmentier, F.J.W.; ~Mi, Y.; ~Maximov, T.C.; ~Dolman, A.J. ~Methane emissions from permafrost thaw lakes limited by lake drainage. \emph{Nat. Clim. Change} ~{\bf 2011}, \emph{1}, 119--123.

\bibitem{sudakov15}
Sudakov, I.;~Vakulenko, S.A.~A mathematical model for a positive permafrost carbon-climate feedback. 
\emph{IMA~J. Appl. Math.}~{\bf 2015}, \emph{80},~811--824.

\bibitem{stepanenkoa11}
Stepanenkova, V.M.; ~Machulskaya, E.E.; ~Glagolev, M.V.; Lykossov, V.N.~Numerical modeling of methane emissions from lakes in the permafrost zone.
\emph{{Izv}. Atmos. Ocean. Phys.}~{\bf 2011}, \emph{47},~252--264.

\bibitem{tan15}
Tan, Z.;~Zhuang, Q. Arctic lakes are continuous methane sources to the atmosphere under warming conditions. \emph{Environ. Res. Lett.} {\bf 2015}, \emph{10}, doi:10.1088/1748-9326/10/5/054016.




















\end{thebibliography}



\end{document}